\documentclass[12pt]{article}
\usepackage{amsmath}
\usepackage{graphicx,psfrag,epsf}
\usepackage{enumerate}
\usepackage{natbib}
\usepackage{url} % not crucial - just used below for the URL 

\usepackage{xcolor}
\usepackage{amsfonts}
\usepackage{amssymb}
\usepackage{amsthm}

\usepackage{bm}
\usepackage[utf8]{inputenc}
\usepackage{soul}

\usepackage{ulem}

\usepackage{pdflscape}
\usepackage{bbm}
\usepackage{subfig}
\usepackage{algorithm}
\usepackage{algorithmic}

\usepackage{blkarray}

\usepackage[colorlinks=true,
            linkcolor=blue,
            urlcolor=blue,
            citecolor=blue]{hyperref}

\usepackage[english]{babel}
\newcommand\mat[1]{\begin{pmatrix}#1\end{pmatrix}}
\usepackage{bigints}

\usepackage{float}
\usepackage{subfig}

\newtheorem{theorem}{Theorem}
\newtheorem{corollary}{Corollary}
\newtheorem{lemma}{Lemma}

\newtheorem{condition}{Condition}

\newtheorem{assumption}{Assumption}

\DeclareMathOperator*{\argmin}{arg\,min}

\DeclareMathOperator*{\argmax}{arg\,max}

\title{Joint Estimation and Inference for Multi-Experiment Networks of High-Dimensional Point Processes}
\author{Xu Wang and Ali Shojaie \\
Department of Biostatistics, University of Washington
}

\begin{document}

\maketitle

\begin{abstract}
Modern high-dimensional point process data, especially those from neuroscience experiments, often involve observations from multiple conditions and/or experiments. Networks of interactions corresponding to these conditions are expected to share many edges, but also exhibit unique, condition-specific ones. However, the degree of similarity among the networks from different conditions is generally unknown. Existing approaches for multivariate point processes do not take these structures into account and do not provide inference for jointly estimated networks. 
To address these needs, we propose a joint estimation procedure for networks of high-dimensional point processes that incorporates easy-to-compute weights in order to data-adaptively encourage similarity between the estimated networks.
%The key feature of our proposed method is the correlation-based weights that encourage similarity between the estimated networks, which facilitate efficient estimation of multi-experiment networks. 
%Extending the results on Hawkes process under a single experiment, we characterize the converge rate of the edge estimation and selection using our proposed method for multi-experiment networks.
%
We also propose a powerful hierarchical multiple testing procedure for edges of all estimated networks, which takes into account the data-driven similarity structure of the multi-experiment networks. Compared to conventional multiple testing procedures, our proposed procedure greatly reduces the number of tests and results in improved power, while tightly controlling the family-wise error rate. Unlike existing procedures, our method is also free of assumptions on dependency  between tests, offers flexibility on $p$-values calculated along the hierarchy, and is robust to misspecification of the hierarchical structure.
%
%by taking advantage of the hierarchical structure of the networks.
%
%Although statistical inference procedure has recently been developed for high-dimensional Hawkes processes under a single experiment, implementing the method directly on networks at all experiments involve a large number of tests. Therefore, we develop a hierarchical testing procedure that reduces the size of tests involved thus increases the testing power by taking advantage of the hierarchical structure of the tests. 
%
We verify our theoretical results via simulation studies and demonstrate the application of the proposed procedure using neuronal spike train data.
\end{abstract}

%%%%%%%%%%%%%%%%%
%% source section files
%%%%%%%%%%%%%%%%%
\newpage 
%%%%%%%%%%%%
\section{Introduction}\label{sec:intro}
%%%%%%%%%%%%
Multivariate point process data have become prevalent in many application areas, from finance and social networks to biology. Of prominent importance are spike train data, containing spiking times of a collection of neurons \citep{Okatan2005}. These data, which have become more abundant thanks to the advent of calcium florescent imaging technology, are increasingly used to learn the latent brain connectivity network and glean insight into how neurons respond to external stimuli. 

The Hawkes process \citep{Hawkes1971} is a popular choice for analyzing multivariate point process data. In this model, the probability of future
events for each component can depend on the entire history of events of other components. As such, the Hawkes process offers a flexible and interpretable framework for investigating the latent network of point processes and is widely used in neuroscience applications
%\textcolor{orange}{
\citep{Brillinger1988,Johnson1996,krumin_reutsky_shoham_2010, Pernice2011, Reynaud2013, TRUCCOLO2016336, LAMBERT20189}.
%}
%\textcolor{red}{[REFS]}. 
 
%
%The latent connectivity structure of these processes can be represented by a probabilistic graphical model
%\citep{lauritzen1996} with a graph or network $G = (V,E)$ whose nodes, $v \in V$, represent components/units in the multivariate point processes and each \textit{directed} edge, $(u \to v) \in E$, indicates that the probability of future events of the target node $v$ depends on the history of the source node $u$. 
%

% high dimension data 
In modern applications, it is common for the number of measured components, e.g., the number of neurons, to be large compared to the observed period, e.g., the duration of neuroscience experiments. The high-dimensional nature of data in such applications poses challenges to learning the connectivity network of a multivariate Hawkes process. \citet{hansen2015} and \citet{Shizhe2017} proposed $\ell_1$-regularized estimation procedures to address this challenge. Recently, \citet{wang2020statistical} developed a high-dimensional inference procedure to characterize the sampling distribution of these estimators and their uncertainty. However, because of the complex dependence structure of the point process data, even regularized estimation and inference procedures require data collected over a long period to achieve satisfactory performance. Unfortunately, available data routinely consist of short stationary segments that may not satisfy these requirements. This is particularly the case in neuroscience applications, where experiments include multiple stimuli that are examined consecutively in order to investigate and contrast how neurons respond to each stimulus. 
%For instance, in their recent study, \citep{Boldingeaat6904} repeatedly apply laser stimulus to a set of neurons at different intensity levels. \textcolor{red}{can we add a figure for their experiment? or add some more details?}
For instance, \citet{Boldingeaat6904} apply 10 laser stimuli to a set of neurons at 8 different intensity levels ranging from 0 to 50 $mW/mm^2$, resulting in 80 experiments in total. 
%\textcolor{red}{can we add a figure for their experiment? or add some more details?}

Neuronal connectivity networks corresponding to different stimuli in a sequence of experiments are expected to share many edges. This shared structure motivates joint estimation of networks from multiple experiments, which could lead to more efficient estimation of common edges. However, the networks from different experiments or conditions are also expected to contain unique edges that are, in fact, of primary scientific interest. For example, distinct neuronal connectivities are found under laser stimulus at different intensity levels \citep{Boldingeaat6904}. Moreover, the degree of (dis)similarity between networks from two experiments depends on the similarity of neuronal responses to the corresponding stimuli, which is generally unknown. 
%Efficient network estimation and inference in such multi-experiment settings requires accounting for similarities between networks while combining data from multiple experiments.
These characteristics highlight the need for joint estimation and inference of multiple networks of high-dimensional point processes while accounting for similarities between networks, a task that is not addressed by existing methods.

%\textcolor{red}{the lit review is very incomplete -- there has been quite a bite of work in this area that needs to be included}
%\textcolor{orange}{
Joint estimation of multiple graphical models is a powerful tool for differential network analysis \citep{shojaie2020} 
and has been considered for independent and Gaussian-distributed data \citep[e.g.][]{Chiquet2011, Guo2011, danaher2011joint, Yajima2014, Zhu2014, Ma2016, Cai2016, Huang2018, wang2020highdimensional}.
Extensions to time- and/or space-varying networks have also been studied \citep[e.g.][]{kolar2010, Qiu2016, Lin2017, Hallac2017, Yang2020}. 
%}
%\textcolor{red}{ALSO OTHER METHODS BLAH BLAH}. 
However, the vast majority of existing approaches are primarily designed for learning \textit{two} networks %   \citet{Yajima2014}
or implicitly assume that networks from multiple experiments are equally similar,  % \citet{Chiquet2011, danaher2011joint, Huang2018, wang2020highdimensional }
%\textcolor{orange}{ 
or that the network similarity is known % e.g. \citep{Zhu2014, Ma2016} 
or implied by the spatial/temporal ordering.    
%these approaches either ignore the underlying (dis)similarities among networks or assume they are known.
%}
On the other hand, the few methods designed for joint estimation of multiple networks \citep{peterson2015, saegusa2016} are specific to Gaussian graphical models and either assume the network edges are independent \citep{peterson2015}, % see Peterson (3.8) where they assume independency on edges
or assume similarities in population means in different conditions reveal similarities among precision matrices \citep{saegusa2016}. 
Moreover, existing procedures do not provide inference for the estimated networks, which is critically important in many scientific applications. 
%
%developed data-driven weights to represent the similarity of sub-populations for Gaussian-distributed data, such weights are calculated from the precision matrix of the joint Gaussian distribution and thus is not applicable to the point process data. Desirable weights also need to be computationally efficient, particularly for networks of large size under many experiments.

Given the paucity of methods for joint analysis of point process data from multiple experiments/conditions, in Section~\ref{sec:estimation} we propose a data-adaptive joint estimation procedure for networks of high-dimensional point processes.  
The proposed approach uses estimates of cross correlations in each condition to obtain a measure of similarities among neuronal connectivity networks. While cross-correlations are widely used by neuroscientists to gain insights into neuronal interaction mechanisms \citep{MAGRANSDEABRIL2018120}, they do not reveal direct interactions between neurons %\textcolor{orange}{
\citep{Tchumatchenko2011, Reid2019}.
%} 
% more lit, \cite{Zanin2013, mehler2020lure}, 
%\textcolor{red}{need a better ref here, ideally something from neuroscience literature}
%
% In the case of Hawkes processes, we take advantage of the one-to-one correspondence between the transfer functions of the Hawkes process and their cross-correlations \citep{Bacry2016} to define data-driven weights that capture the similarities among neuronal connectivity networks defined based on transfer functions. 
%
However, they can be easily computed, even in high dimensions. We also show that they provide useful information about the overall similarities among neuronal connectivity networks. In particular, we show that cross-correlations can be used to define  data-driven weights for joint estimation of multiple networks.
% especially when used to define data-driven weights for joint estimation of multiple networks. 
By encouraging similarity among estimated networks, such data-driven weights offer superior finite-sample performance in selecting the edges.
% add theoretical results in est. consistency. 
Extending previous work under a single experiment \citep{Shizhe2017}, in Section~\ref{sec:theory} we establish a unified non-asymptotic convergence rate for edge estimation in multiple networks of generalized Hawkes processes. %without assuming that the true weights are known
%\textcolor{red}{we probably need to update this sentence based on the results}. 
Unlike previous theoretical analysis, our result implies a faster convergence rate using the joint estimation approach compared with estimating networks separately under each experiment. 
%\as{does it really imply a faster `rate'?} \sw{as mentioned later, faster rate needs $M$ goes inf?}
More specifically, while our method does not assume the correctness of the weights to achieve the estimation convergence, we achieve a lower estimation error when the true weights are known. %\as{where do we show this?}\sw{sw: Yes, see \eqref{eq:bound_D} in the proof of Theorem~1. will clarify when discuss Theorem~1 result later.}

% talk about inference
To address the need for efficient inference procedures, in Section~\ref{sec:test}, we develop a hierarchical testing procedure for simultaneous inference on edges of all estimated networks. 
% Evaluating the uncertainty of the network estimates is critical in scientific applications. 
While statistical inference for a single high-dimensional Hawkes processes has been recently addressed \citep{wang2020statistical}, implementing such an approach directly on all estimated networks would amount to a large number of tests. As a result, the testing power can diminish after adjusting for multiple comparisons, especially when the number of experiments is large. This is particularly the case in neuroscience applications, such as that in \citet{Boldingeaat6904} with 80 experiments. 
Moreover, the tests associated with network edges under each experiment have complex dependencies. %\textcolor{red}{I'm confused --- do you mean the tests for the same edge in different conditions? SW: no, I mean the $p$ tests for edges under the same experiments. Because we repeatedly use data from the same experiment to calculate the test statistics for each of the tests, those test statistics are correlated.} 
Our proposed inference framework mitigates these challenges by testing hypotheses along the hierarchical structure of network similarities inferred from  cross-correlations. 
%To plug the gap, we propose a powerful hierarchical testing procedure on the edges of all networks, where the hierarchical structure between the networks is guided by the cross-correlation based weights. 
By taking advantage of this hierarchical structure, our procedure greatly reduces the number of required tests, which in turn results in increased power while tightly controlling the family-wise error rate (FWER). While motivated by neuroscience applications, the framework is also general and can be applied to testing hypotheses corresponding to joint estimation across multiple conditions. 

%\textcolor{orange}{
Unlike existing hierarchical testing procedures that rely on specific assumptions on the dependency between the tests \citep{Yekutieli2008,lynch2016procedures} or particular $p$-value calculation along the hierarchy \citep{Bogomolov2020}, our method is free of such assumptions and can incorporate $p$-values flexibly calculated from any valid test. Moreover, as we show in Section~\ref{sec:test}, for large and sparse networks, our inference procedure is robust to misspecification of the similarity weights, which had not been previously addressed.
%}
The advantages of our estimation and inference procedures are illustrated by analyzing simulated and real data in Sections~\ref{sec:sims} and \ref{sec:data}, respectively. 

The implementation of our proposed joint estimation and inference procedure in \texttt{python} is available at \url{https://github.com/stevenwang/NeuroNetLearn}.

%
%Unlike the existing hierarchical testing methods \citep[e.g.][]{Yekutieli2008,lynch2016procedures} that control the FWER over all the tests involved, our proposed method controls the error rate among the hypothesis associated only with the leaves of the dendrogram, which is exactly the set of hypotheses of our interest. In addition, our method is free of the dependency assumption over the tests and is robust to mis-specification of the hierarchical structure of the networks for large and sparse networks.

%The rest of this paper is organized as follows. Section~\ref{sec:hawkes} introduces the generalized Hawkes process and reviews its basic properties. Our proposed joint estimation procedure
%is outlined in Section~\ref{sec:estimation}. In Section~\ref{sec:theory}, we present theoretical results that guarantee the edge selection using our proposed procedure. The construction of our inference procedure is presented in Section~\ref{sec:test}. We investigate the properties of the proposed estimation and inference procedure using simulations in Section~\ref{sec:sims} and illustrate
%its utility in neuroscience applications in Section~\ref{sec:data}. We add computation details and proofs of the main theorems in the Appendix. 

% think unconditional and conditional network meaning
%%%%%%%%%%%%%
%%%%%%%%%%%%%
\section{Background: The Hawkes Process}\label{sec:hawkes}
%%%%%%%%%%%%%
Let $\{t_k\}_{k\in \mathbb{Z}}$ be a sequence of real-valued random variables,
taking values in $[0, T]$, with $t_{k+1} > t_k$ and $t_1 \ge 0$.
Here, time $t = 0$ is a reference point in time, e.g., the start of an
experiment, and $T$ is the duration of the experiment. A simple point process
$N$ on $\mathbb{R}$ is defined as a family $\{ N(A) \}_{ A \in
	\mathcal{B}(\mathbb{R}) }$, where $\mathcal{B}(\mathbb{R})$ denotes the Borel
$\sigma$-field of the real line and $N(A) = \sum_k \mathbf{1}_{\{t_k \in A\}} $.
The process $N$ is essentially a simple counting process with isolated jumps of
unit height that occur at $\{t_k\}_{k\in \mathbb{Z}}$. We write $ N([t, t + dt)
)$ as $dN(t)$, where $dt$ denotes an arbitrarily small increment of $t$.

Let $\mathbf{N}$ be a $p$-variate counting process $\mathbf{N} \equiv \{
N_i\}_{i\in \{1,\dots, p \}}$, where, as above, $N_i$ satisfies $N_i(A) = \sum_k
\mathbf{1}_{\{t_{ik} \in A\}}$ for $A \in \mathcal{B}(\mathbb{R})$ with $\{
t_{i1}, t_{i2}, \dots \}$ denoting the event times of $N_i$.  Let
$\mathcal{H}_t$ be the history of $\mathbf{N}$ prior to time $t$.  The intensity
process $\{ \lambda_{1}(t), \dots, \lambda_{p}(t) \}$ is a $p$-variate
$\mathcal{H}_t$-predictable process, defined as
\begin{align} \label{eq:intensity_prob}
\lambda_i (t) dt & = \mathbb{P}(dN_{i}(t)=1 \mid \mathcal{H}_t )  .
\end{align}
The intensity function for the Hawkes process \citep{Hawkes1971}, takes the form
\begin{align}
\lambda_{i}(t)
&=  g_i\left( \mu_{i} + \sum_{j=1}^{p} \left(\omega_{ij} *  dN_{j} \right )(t) \right)
\label{eq:hawkes}  ,
\end{align}
where
\begin{align}
\left( \omega_{ij} *  dN_{j}  \right)(t)  =
\int_0^{t-}  \omega_{ij}(t-s)   dN_{j} (s)
= \sum_{ k: t_{jk} < t }   \omega_{ij}( t- t_{jk}) . \label{eq:transfer_function}
\end{align}
Here, $\mu_{i}$ is the background intensity of unit $i$ and
$\omega_{ij}(\cdot): \mathbb{R}^+ \mapsto \mathbb{R}$ is the
\textit{transfer function}. In particular, $\omega_{ij}( t - t_{jk}) $
represents the influence from the $k$th event of unit $j$ on the intensity
of unit $i$ at time $t$.
If the link function $g_i(\cdot)$ is non-linear, then $\lambda_i(\cdot)$ is the intensity of non-linear Hawkes process \citep{Bremaud1996}. We refer to the class of Hawkes
processes that allows for non-linear link functions and negative transfer functions as the \textit{generalized Hawkes process} \citep{Shizhe2017}.

Motivated by applications in neuroscience \citep{Linderman2014, MAGRANSDEABRIL2018120}, we consider a parametric transfer function
$\omega_{ij}(\cdot)$ of the form
\begin{equation} \label{eq:omega}
\omega_{ij} (t)    =   \beta_{ij} \kappa_{j}(t) 
\end{equation}
with a \textit{transition kernel}
$\kappa_j(\cdot): \mathbb{R}^+ \rightarrow \mathbb{R}$ that captures the decay of the dependence on past events. This leads to $ \left( \omega_{ij} * dN_{j}  \right)(t) = \beta_{ij} x_{j}(t)$, where the \textit{integrated stochastic process}
\begin{equation}\label{eq:design_column_xt}
x_{j}(t)  = \int_0^{t-} \kappa_{j}(t-s) dN_{j}(s) 
\end{equation}
summarizes the entire history of unit $j$ of the multivariate Hawkes processes. A commonly used example is the exponential transition kernel, $\kappa_{j}(t) = e^{-t}$ \citep{Bacry2015}.

In this formulation, the \textit{connectivity coefficient} of the underlying network, $\beta_{ij}$, represents the strength of the dependence of unit $i$'s intensity on unit $j$'s past events. A positive $\beta_{ij}$, which implies that past events of
unit $j$ \textit{excite} future events of unit $i$, is often considered in the
literature \citep[see, e.g.,][]{Bacry2015, Etesami2016}. We also allow for negative $\beta_{ij}$ values to represent
\textit{inhibitory} effect of one unit's past events on another
\citep{Shizhe2017,costa2018}, expected in neuroscience applications \citep{Purves2001}. 

Denoting
$\bm{x}(t)= ( x_1(t), \dots, x_p(t) )^\top \in \mathbb{R}^{p}$ and
$\bm{\beta}_i = (\beta_{i1}, \dots, \beta_{ip})^\top \in
\mathbb{R}^{p}$, we can write
\begin{align}
\lambda_{i}(t)
&=  g_i\left(  \mu_{i} + \bm{x}^\top(t)\bm{\beta}_i  \right) \label{eq:hawkes_para_transfer} .
\end{align}
\section{Networks of Multi-Experiment Hawkes Processes}\label{sec:estimation}
%%%%%%%%%%%%%
\subsection{Joint Estimation via Regularization}
Given point process data from $M$ experiments, let $N_i^{(m)}$ be the point process of unit $i$ under experiment $m$ defined on $[0, T_m]$, where $m\in\{1,\dots,M\}$. We use experiment-specific notations for the corresponding intensity function $\lambda^{(m)}_i(\cdot)$, link function $g_i^{(m)}(\cdot)$, background rate $\mu^{(m)}_i$, connectivity coefficient $\beta_{ij}^{(m)}$, transfer function $\omega^{(m)}_{ij}(\cdot)$, transition kernel function $\kappa^{(m)}_j(\cdot)$, and integrated process $x_j^{(m)}(\cdot)$. We also denote the entire model parameter for unit $i$ at experiment $m$ as $\bm{\theta}_i^{(m)} = \left( \mu_i^{(m)},  \big( \bm{\beta}^{(m)}_i \big)^\top \right)^\top$
% $\bm{\theta}_i^{(m)} = \mat{ \mu_i^{(m)}\\ \bm{\beta}^{(m)}_i } $
, where $\bm{\beta}^{(m)}_i= \mat{\beta^{(m)}_{i1} & \dots & \beta^{(m)}_{ip} }^\top$; and let $\bm{\theta}_i = \mat{ \big( \bm{\theta}_i^{(1)} \big)^\top & \dots & \big( \bm{\theta}_i^{(M)} \big)^\top }^\top $, 
$\bm{x}^{(m)}(t) = \mat{ x_1^{(m)}(t) & \dots & x_p^{(m)}(t) }^\top$, and $T = \sum_{m=1}^M T_m$.

Throughout the paper, we assume that the generalized Hawkes process \eqref{eq:hawkes_para_transfer} from each experiment 
is \textit{stationary}, meaning that for all units
$i\in\{ 1,\dots, p\}$, the spontaneous rates $\mu^{(m)}_i$ and strengths of
transition $\bm{\beta}^{(m)}_i$ are constant over the time range $[0,T_m]$ under each experiment $m\in\{ 1,\dots, M\}$ \citep{Bremaud1996}. This assumption is often satisfied in neuroscience applications, 
%\textcolor{orange}{
where a ``white noise" period is included between consecutive experiments---for instance, \citet{Boldingeaat6904} turn off the laser for one second between consecutive stimuli.
%}

%\textcolor{red}{say something about the white noise periods between each two experiments...}

Let $\ell(\cdot,\cdot): \mathbb{R} \times \mathbb{R} \rightarrow \mathbb{R}^+$ be a twice continuously differentiable loss function. 
%We take $\ell(\cdot, \cdot)$ as the least square loss if the link function is linear \citep{hansen2015, wang2020statistical} and negative likelihood if the link function is non-linear.
To compactly define our penalized estimator, we use \eqref{eq:hawkes_para_transfer} to define the expectation of the observed %point process 
outcome at $[t,t+dt)$ for unit $i$ under experiment $m$ as 
$$ f_{\bm{\theta}^{(m)}_i }\left (\bm{x}^{(m)}(t) \right) \equiv \lambda^{(m)}_i(t) dt =
g^{(m)}_i \left( \mu^{(m)}_i
+  \left( \bm{x}^{(m)}(t) \right)^\top \bm{\beta}^{(m)}_i \right) dt. 
$$ 
Our proposed joint estimation procedure is then characterized by the following optimization problem:
\begin{align}
\left\{ \widehat{\bm{\theta} }_i  \right \}_{1\le i \le p} = 
\argmin_{
	\substack{
		\bm{\theta}^{(m)}_i 
		\in \mathbb{R}^{ p+1}\\ 1\le i \le p , 1\le m \le M 
	} 
}
\sum_{i=1}^p 
\left\{
\frac{1}{ T }
\sum_{m=1}^M
\int_{0}^{T_m}
\ell \left (   dN^{(m)}_i(t), f_{\bm{\theta}^{(m)}_i }(\bm{x}^{(m)}(t)) \right )   + 
\mathcal{P}\left ( \left \{ \bm{\beta}_i^{(m)} \right \}_{m=1}^M \right )  
\right\},
\label{eq:optimization}
% .
\end{align}
where to achieve both network sparsity and similarity, we consider the penalty %\textcolor{red}{we have too many lambdas -- can we use $\rho_1$ and $\rho_2$ for tuning and replace $\rho_c$ and $\rho_r$ with $\tau_c$ and $\tau_r$? } 
%\sw{updated $\lambda_1,\lambda_2$ to $\rho_1,\rho_2$.}
\begin{align}
\mathcal{P}\left ( \left  \{ \bm{\beta}_i^{(m)} \right \}_{m=1}^M \right ) 
= 
\underbrace{ \rho_1 \sum_{m=1}^M  \left \lVert \bm{\beta}^{(m)}_i \right \rVert_1}_{
	\text{sparsity penalty}
}
+ \underbrace{ \rho_2 \sum\limits_{1\le m < m' \le M} w_{m,m'} \sum \limits_{1\le i \le p} \left \lVert  \bm{\beta}^{(m)}_{i} - \bm{\beta}^{(m')}_{i}  \right \rVert_1
}_{\text{fusion penalty}}.
\label{eq:penalty}
\end{align}
The sparsity penalty encourages a sparse structure of the estimated networks (i.e. few non-zero $\beta^{(m)}_{ij}$, for $1\le m\le M$). The fusion penalty \citep{TibsFused2005} encourages similarity in the estimated networks. 
The key feature of our penalty is that the extent of fusion is governed by the {\it data-driven weights} $w_{m,m'} \in [0,1]$ for $1\le m,m'\le M$. A larger weight represents more similar networks in two experiments $m$ and $m'$.

\subsection{Data-Driven Similarity Weights}
%% network similarity and weight 
To construct the similarity weight between networks, $w_{m,m'}$, we start by measuring the similarity between two matrices, $A=\left\{ a_{ij} \right\}, A' =\left\{ a'_{ij} \right\} \in \mathbb{R}^{p\times p}$, by
\begin{align}\label{eq:mat_similarity}
    d(A, A') \equiv \sum_{1\le i,j \le p} \mathbf{1} \left( a_{ij} a'_{ij} > 0 \right)  .
\end{align}
In the context of network analysis, $d(\cdot, \cdot)$ measures the similarity of two networks according to their connectivity structures. Specifically, consider the $p \times p$ adjacency matrices $\mathsf{B}^{(m)} =\Big\{ \mathbf{1}_{ \beta^{(m)}_{ij} \ne 0 } \Big\} $ and $\mathsf{B}^{(m')} =\Big\{ \mathbf{1}_{ \beta^{(m')}_{ij} \ne 0 } \Big\} $ for networks under condition $m$ and $m'$. The network similarity is then given by 
%\textcolor{red}{please note that I changed the notation of matrices here}
\begin{align}    \label{def:oracle_dist}
d^o_{m,m'} \equiv d\left(\mathsf{B}^{(m)}, \mathsf{B}^{(m')}\right),
\end{align}
which is well-defined as $p^2-d^o_{m,m'} $ is the \textit{Hamming distance} between graphs $m$ and $m'$, measuring the difference in their connectivity structures. 
%\textcolor{red}{what if the betas have opposite signs but are both nonzero? then it won't be the hamming distance? would it be easier to define this based on the 0-1 adj matrix, which is what we need below?}
%\sw{That makes sense. I can define $\mathsf{B}^{(m)} =\left\{ \mathbf{1}_{ \beta^{(m)}_{ij} \ne 0 } \right\} $, so that the entry of the matrix indicate adjacency.}
%
We call $d^o_{m,m'}$ the \textit{oracle network similarity} because, in practice, the true connectivity matrices are unknown. As a surrogate, we propose a measure based on cross-covariances \citep{Hawkes1971}: 
 \begin{align}    \label{def:cv_dist}
d^c_{m,m'} \equiv d(V^{(m)}, V^{(m')}), 
\end{align}
where $V^{(m)}=\{ V^{(m)}_{ij} \}$ and $ V^{(m')}=\{ V^{(m')}_{ij} \}$ are cross-covariance matrices for networks under condition $m$ and $m'$.
While they only represent marginal temporal associations between component processes, cross-covariances can be easily computed, even in high-dimensional settings. Moreover, they tend to capture overall network similarities  in different conditions. In fact, for mutually-exciting networks, there exists a one-to-one mapping between the cross-covariance matrix and the connectivity matrix  \citep[see][Theorem~1]{Bacry2016}, and the set of edges represented by the non-zero cross-covariance is a super set of the true edge set \citep[see][Lemma~1]{ShizheEJS2017}. In practice, we estimate the cross-covariance matrix using its empirical estimate, $\widehat{V}^{(m)}_{ij} =\left\{ \widehat{V}^{(m)}_{ij} \right\}$ \citep{ShizheEJS2017}. Given a pre-specified threshold $\kappa > 0$, the thresholded sample cross-covariance matrix is given by
\begin{align*}
  \widetilde{V}^{(m)}_{ij} = \widehat{V}^{(m)}_{ij}\mathbbm{1}\left( \big | \widehat{V}^{(m)}_{ij} \big |  > \kappa   \right) .
\end{align*}
Our proposed \textit{empirical network similarity} is then defined as  
\begin{align}\label{eq:empirical_similarity}
    d^{e}_{  m, m' } \equiv d\left(\widetilde{V}^{(m)}_{ij}  , \widetilde{V}^{(m')}_{ij} \right). 
\end{align}
By the consistency of the sample cross-covariance estimator  \citep[see][Corollary~1]{Shizhe2017}, it follows that the empirical network similarity consistently estimates the similarity based on the true cross-covariance if the true cross-covariances are larger in magnitude than the minimum signal strength, $\kappa$. 
%, in order to distinguish them from $0$, given the amount of information observed. 
In practice, to put these sample cross-covariances in a comparable range, we transform the covariance values into $z$-scores using Fisher's transformation and obtain the corresponding $p$-values. We claim an edge in this cross-covariance network if the $p$-value is below a pre-specified threshold, e.g. 0.1. The threshold can also be chosen to achieve a certain level of sparsity in the network \citep{ShizheEJS2017}. 
The \textit{similarity weights} used in our algorithm are obtained by normalizing the empirical similarity measure by their total so that the weights are always between 0 and 1; that is,
\begin{align}\label{eq:empirical_weights}
    w_{m,m'} \equiv  \frac{ d^{e}_{  m, m' }   }{  
    \sum_{1\le k' \ne k \le M}  d^{e}_{ k, k' }  } \in [0,1],  
\end{align}
for $1\le m \ne  m' \le M$.

\subsection{Computation and Tuning}
Fusion penalty has been used in previous work but with a natural ordering of the conditions defined according to location or network structure \citep{TangLu2016,Wang2016}. Given that no ordering necessarily exists between experiments, we adopt a different computation strategy based on the smoothing proximal gradient descent algorithm \citep{ChenSPGD2012} to solve \eqref{eq:optimization}. Implementation details are given in Appendix~\ref{sec:algo}.

The optimization problem in \eqref{eq:optimization} involves two tuning parameters that are used to control the sparsity and the similarity of the networks between experiments. We use an eBIC-type criterion \citep{chenchen2008,cai2020latent} to select these parameters. Let $ \widehat{ \bm{\Theta}}_{\rho_1,\rho_2} =\left \{ \widehat{ \bm{\theta} }^{(m)} (\rho_1,\rho_2)  \right \}_{m=1}^M$ be the model parameter estimates using tuning parameters $(\rho_1,\rho_2)$. Then, 
\begin{align*}
 \mathrm{eBIC}\left(\widehat{ \bm{\Theta}}_{\rho_1,\rho_2} \right)  =
 \sum_{m=1}^M \sum_{i=1}^p   
 \left\{ 
 2 \ell \left(   dN^{(m)}_i(t), f_{\bm{\theta}^{(m)}_i(\rho_1,\rho_2)  }\left(\bm{x}^{(m)}(t)\right) \right)
 +  s^{(m)}_i \log T_m
 + 2 \gamma \log { p \choose s^{(m)}_i } 
 \right \} ,
\end{align*}
where $s_i^{(m)} =  \left \lVert  \widehat{ \bm{\beta}}^{(m)}_i(\rho_1,\rho_2)   \right \rVert_0$
and ${ p \choose s^{(m)}_i }$ is the binomial coefficient. 
%We use $\gamma =0.1 $ in our empirical studies. 
% comment on eBIC convergence rate for point process data is unknown, thus in practice, we recommend also select tuning parameters to the best match with domain knowledge.

For ease of presentation in later sections, let $d_{m,m'} = w_{m,m'} (\bm{e}^\top_{m} - \bm{e}^\top_{m'}) \otimes I_0 \in \mathbb{R}^{(p+1)\times (p+1)M}$ and denote 
$D = \left(d_{1,2}, \ldots, d_{m,m'}, \ldots, d_{M-1, M} \right)^\top \in \mathbb{R}^{ {M \choose 2} (p+1) \times (p+1)M}$,
% $D = \mat{ d_{1,2} \\
% 	\dots \\
% 	d_{m,m'} \\
% 	\dots \\ d_{M-1, M}  } \in \mathbb{R}^{ {M \choose 2} (p+1) \times (p+1)M}$,
% and $d_{m,m'} = w_{m,m'} (\bm{e}^\top_{m} - \bm{e}^\top_{m'}) \otimes I_0 \in \mathbb{R}^{(p+1)\times (p+1)M} $, 
where $\bm{e}_m$ is the canonical basis in $\mathbb{R}^M$, $I_0 = \mat{0 & 0 \\ 	0 & I_p} \in \mathbb{R}^{(p+1)\times (p+1)}$,  
and $I_p \in \mathbb{R}^{p\times p}$ is the identity matrix. Then, the fusion penalty can be written compactly as
\begin{align}\label{eqn:compactfusion}
  \sum_{1\le m < m' \le M} w_{m,m'}  \sum_{1\le i \le p} \left \lVert \bm{\beta}^{(m)}_i - \bm{\beta}^{(m')}_i \right \rVert_1
=   \left \lVert D \bm{\theta} \right \rVert_1. 
\end{align}

%%%%%%%%%%%%%
\section{Network Estimation Consistency}\label{sec:theory}
%%%%%%%%%%%%%
In this section we establish consistent parameter estimation and recovery of the latent networks over multiple experiments. Proofs for the results in this section are given in Appendix~\ref{sec:proofs}.

We start by stating our assumptions. Throughout, we denote the maximum and minimum eigenvalues of a square matrix $A$ as $\Lambda_{\max}(A)$ and $\Lambda_{\min}(A)$, respectively.

\begin{assumption}\label{assumption1}
	The link functions, $g^{(m)}_i(\cdot)$, are first-order differentiable such that $| \nabla g^{(m)}_i(\cdot)|\le \alpha^{(m)}_i$, for $1 \le i \le p, 1\le m \le M$. Further, let $\Omega^{(m)}$ be a $p\times p$ matrix whose entries are $\Omega^{(m)}_{ij} =  \alpha^{(m)}_i \int_0^{\infty} |\omega^{(m)}_{ij}(\Delta)| d\Delta$, for $1 \le i,j \le p, 1\le m \le M$. Then, there exists a constant 
	$\gamma_{\Omega}$ such that $\Lambda_{\max} ( (\Omega^{(m)} )^\top \Omega^{(m)} ) \le \gamma^2_{\Omega}  < 1 $, for $1\le m \le M$.
\end{assumption}

Assumption~\ref{assumption1} is necessary for stationarity of a Hawkes process under each specific experiment \citep{Shizhe2017}. The constant $\gamma_{\Omega}$ does not depend on the dimension $p$.  For any fixed $p$, \citet{Bremaud1996} show that given this assumption the intensity process of the form \eqref{eq:hawkes} is stable in distribution and, thus, a stationary process exists. 
Since the connectivity coefficients of interest are ill-defined without stationarity, this assumption provides the necessary context for our joint estimation framework.

\begin{assumption}\label{assumption2}
	There exists  $\lambda_{\max}$ such that
	$
	 \lambda^{(m)}_{i}(t) \le \lambda_{\max} < \infty ,
	\quad t\in [0, T_m]
	$
	for all $i=1,\dots, p$ and $m=1,\dots,M$.
\end{assumption}
Assumption~\ref{assumption2} requires that intensities are upper bounded. Similar assumptions are commonly considered in the analysis of multivariate Hawkes processes \citep{hansen2015,costa2018,Shizhe2017, cai2020latent}.
% to allow the min-eigen value condition, we need E lambda_i(t) > 0, this can be done assumption the background intensity g(u_i) > 0. For linear model, we can assume u_i > 0; for non-linear link, like exp(), this is always satisfied as long as u_i is not -infity.

\begin{assumption}\label{assumption3}
	The transition kernel $\kappa^{(m)}_i(t)$ is bounded and integrable over $[0,T_m]$, for $1\le i \le p$ and $1\le m \le M$.
\end{assumption}
Assumption~\ref{assumption3} implies that the integrated process $x_i^{(m)}(t)$
in \eqref{eq:design_column_xt} is bounded. 
Together, Assumptions~\ref{assumption2} and \ref{assumption3} imply that the model parameters
are bounded, which is often required in time-series settings \citep{Safikhani2017JointSB}.

\begin{assumption}\label{assumption4}
	There exists constants $\tau_r \in (0,1)$ and $0< \tau_c < \infty $ such that 
	%\as{maybe change rohs to tau?}
	%\sw{updated $\rho_c, \rho_r$ to $\tau_c, \tau_r$}
	\begin{align*}
	\max_{1\le i \le p}  \sum_{j=1}^p \Omega^{(m)}_{ij} \le  \tau_r
	\qquad\text{and}\qquad
	\max_{1\le j \le p}   \sum_{i=1}^p \Omega^{(m)}_{ij} \le \tau_c ,
	\end{align*}
	for $m=1,\dots,M$.
\end{assumption}
Assumption~\ref{assumption4} requires maximum in- and out- intensity flows to be bounded, which helps
in bounding the eigenvalues of the cross-covariance of $\bm{x}^{(m)}(t)$ \citep{wang2020statistical}. A similar assumption is also considered by \citet{Basu2015} in the context of VAR models. 
%
%Assumption~\ref{assumption4} requires maximum in- and out- intensity flows to be
%bounded. A similar assumption is also considered by \citet{Basu2015} in
%the context of VAR models. This assumption helps in bounding the eigenvalues of the cross-covariance of $\bm{x}^{(m)}(t)$. 
%Let
%$
%\textrm{Q} = \mat{ \textrm{Q}^{(1)} &         &          & \\ 
%	& \textrm{Q}^{(2)} &          & \\  
%	&         &  \dots   & \\  
%	&         &          & \textrm{Q}^{(M)} 
%} ,
%$
%where $\textrm{Q}^{(m)}= \int_0^{T_m}   \mat{1 \\ \bm{x}^{(m)}(t) } 
%\mat{1 &  \bm{x}^{(m)}(t)} dt$, and  
% $\gamma = \min_{1\le m \le M} \Lambda_{\min}\left( \textrm{Q}^{(m)}  \right)$. 
%%In the literature of penalized regression \citep{bickel2009,Chichignoud2014}, 
%%$\gamma$ is referred as restricted eigen-value (RE). 
%\citet{wang2020statistical} show that under Assumption~\ref{assumption1}--~\ref{assumption4}, $\Lambda_{\min}\left( \textrm{Q}^{(m)}  \right)$ is strictly positive with high probability. 

Define the set of active indices as $S^{(m)}_i= \{j:  \beta^{(m)}_{ij}\ne 0 , 1\le j \le p \}$, and let $d^{(m)}_i = |S^{(m)}_i|$, $d^* \equiv \max_{ 1\le m\le M, 1\le i \le p} d_i^{(m)} $, and $S_i = \bigcup_{m=1}^M S_i^{(m)}$. 
Also denote the set of dissimilar experiment indices as $\widetilde{S}_i= \left \{ (j,m): \beta_{ij}^{(m)}\ne \beta_{ij}^{(m')},  \exists \, m'\ne m \in \{1 ,\dots, M\}, \text{ for } 1\le j \le p \right \}$. Define the dissimilarity index $ r^* \equiv \max_{1\le i \le p} |\widetilde{S}_i|$, where $r^*=0$ for $M=1$. With a slight abuse of notation, we write $m\in \widetilde{S}_i$ if $\exists \, j$ such that $(j,m) \in \widetilde{S}_i$.
%$\widetilde{S}_{ij} = \left \{ (j,m): \beta_{ij}^{(m)}\ne \beta_{ij}^{(m')},  \exists  m'\ne m \in \{1 ,\dots, M\}  \right \}$ and $\widetilde{S}_i = \bigcup_{j=1}^p \widetilde{S}_{ij}$.
%
%It follows that $|S_i|\le Md^*$ and thus $ r^*  \le Md^*$. 
% \sw{
% Let $A_i = S_i \cap \widetilde{S}_i$. It follows that $|A_i| \le d^* + r^* $. 
% }
%
In addition, for $\widehat{\bm{\theta} }_i^{(m)}$ defined in \eqref{eq:optimization}, let $\Delta_i^{(m)} = \widehat{\bm{\theta} }_i^{(m)}  - \bm{\theta}_i^{(m)}$ 
and $\Delta_i = \left ( \left ( \Delta_i^{(1)}  \right )^\top, \dots, \left ( \Delta_i^{(M)} \right)^\top \right )^\top \in \mathbb{R}^{(p+1)M} $. With these notations, $\Delta_{S_i}$ and $ \Delta_{ \widetilde{S}_i } $ are vectors that collect the estimation error on $\beta_{ij}^{(m)}$ that are non-zero, and those varying over the experiments, respectively.%} 

Let 
$$
\mathcal{C} =
\left \{ 
\Delta \in R^{M(p+1)} : 
 \frac{1}{\sqrt{M}} \lVert \Delta_{S_i^c } \rVert_1 + 2 \lVert  D_{., \widetilde{S}_i^c} \Delta_{ \widetilde{S}_i^c }\rVert_1
\le \frac{3}{\sqrt{M}} \lVert \Delta_{S_i} \rVert_1 
+ 2  \lVert  D_{.,\widetilde{S}_i } \Delta_{ \widetilde{S}_i } \rVert_1 
\right \},
$$ 
%Here $D = (\bm{d}_1,\dots, \bm{d}_{M-1})^\top \in \mathbb{R}^{ {M \choose 2}\times (p+1)M }$,  where $\bm{d}_m = (\bm{d}_{m,m+1},\dots,\bm{d}_{m,M})^\top$ and $\bm{d}_{m,m'} = w_{m,m'} (\bm{e}_m - \bm{e}_{m'}) \otimes (0,\mathbf{1}_{p}^\top) \in \mathbb{R}^{(p+1)M}$. 
where $D$ was defined in \eqref{eqn:compactfusion} and $ D_{.,\widetilde{S}_i } $, $ D_{.,\widetilde{S}^c_i } $ are columns of $D$ corresponding to index sets $  \widetilde{S}_i  $, $ \widetilde{S}^c_i  $, respectively. 
Next, we introduce two conditions that are required on $\ell(  t;\bm{\theta}^{(m)}_i ) \equiv  \ell \left (   dN_i^{(m)}(t), f_{\bm{\theta}_i}^{(m)} (\bm{x}^{(m)}(t)) \right )$. 

\begin{condition}\label{def:RSC}
	There exist constants $\eta, c, C >0 $ such that, for $1\le i \le p$,
	\begin{align*}
	\mathbb{P}\left(
	\min_{\Delta \in \mathcal{C}} 
	\frac{1}{T}\sum_{m=1}^M  
	\Delta_i^\top \left( \int_0^{T_m}  \nabla^2\ell(t;\bm{\theta}_i^{(m)}) \right) \Delta_i   \ge  \eta \lVert \Delta \rVert^2_2 
	\right) \ge 1-c p^2  \sum_{m=1}^M T_m \exp(- C T_m^{1/5}) .
	\end{align*}
\end{condition}

The first condition is known as the restricted strong convexity (RSC) \citep{Negahban2012}. 
The constraint set, $\mathcal{C}$, is constructed specifically for the penalty in \eqref{eq:penalty}, which is geometrically decomposable  \citep{lee2015}. This construction links the estimation error bound to both the sparsity and dissimilarity of the multi-experiment networks. In  Corollary~\ref{corollary1}, we show that this condition is satisfied for linear Hawkes process or generalized Hawkes process with exponential-link under Assumption~\ref{assumption2}.

\begin{condition}\label{def:tail_bound}
There exist $c,C >0$ such that, for $1\le i \le p$,
\begin{align*}
\mathbb{P}\left( 
\left \lVert   
- \frac{1}{T} \sum_{m=1}^M \int_0^{T_m} \nabla \ell(  t;\bm{\theta}^{(m)}_i )  \right \rVert_\infty 
\le  
C T^{-2/5}  
 \right) \ge 1- c pM \exp(- M^{-1} T^{1/5} ) .
\end{align*}
\end{condition}
% Condition~\ref{def:tail_bound} is satisfied with common loss functions such as the least square loss and the negative-likelihood loss (see Corollary~\ref{corollary1}).
% \sw{
% This condition is a technical requirement in showing estimation consistency in the penalized regression problem (see Theorem~\ref{theorem1}). Such condition is satisfied with common loss functions such as the least square loss and the negative-likelihood loss (see Corollary~\ref{corollary1}). }

The second condition is a technical condition needed to establish estimation consistency of penalized regression (see Theorem~\ref{theorem1}). 
% and is satisfied for common loss functions such as the least square loss and the negative-likelihood loss (see Corollary~\ref{corollary1}).
In Corollary~\ref{corollary1}, we also show that this condition is satisfied for common loss functions such as the least square loss and the negative-likelihood loss. 
The lower bound in this condition could be potentially improved to 
$1-c p^2  T \exp(- C T^{1/5})$. However, this requires examination of the minimum eigenvalues of the Hessian matrix---i.e.,$\frac{1}{T}\int_0^{T}  \nabla^2\ell(t;\bm{\theta}_i^{(m)})  $--- %\as{what matrix?} 
for a non-stationary process over all experiments of duration $T= \sum_{m=1}^M T_m$
%\sw{---i.e., considering the multi-experiment processes as one process over $T= \sum_{m=1}^M T_m$---} 
where the process in each experiment is stationary.

% \sw{
% Note that we may consider a tighter probabilistic bound as $1-c p^2  T \exp(- C T^{1/5}) $. However, validating such condition involves a challenging examination on the min-eigenvalue condition for a non-stationary process that has its stationary segment under each experiment.}

%
% Condition~\ref{def:RSC} is referred as the restrict strong convexity (RSC) \citep{Negahban2012}. %This condition is satisfied for linear Hawkes process or generalized Hawkes process with exponential-link under Assumption~\ref{assumption2}.
% %
% The constraint set, $\mathcal{C}$, is constructed specifically to the penalty in \eqref{eq:penalty} which is referred as geometrically decomposable in the literature of regularized $M$-estimator \citep{lee2015}. Such construction links the estimation error bound to both the sparsity and dissimilarity of the multi-experiment networks. 

\begin{theorem}
	\label{theorem1}
	Assume the $p$-variate Hawkes processes for all $M$ experiments---with each component process has its intensity function defined in \eqref{eq:hawkes}---satisfy Assumptions~\ref{assumption1}--~\ref{assumption4}. In addition, suppose  Conditions~\ref{def:RSC} and~\ref{def:tail_bound} are met and $\log p = o(\min T^{1/5}_m)$ and $ (d^* \vee r^* )^{1/4} = o(T^{1/5})$, where $T= \sum_{m=1}^M T_m$. 
	Then, taking $\sqrt{M}\lambda_1 = \lambda_2 =  O(T^{-2/5})$,
	\begin{align}
	\lVert \Delta_i \rVert_2  = O_p\left(  \frac{1}{\eta }
	T^{-2/5}\left( 3\sqrt{d^*} + 2\phi_{\widetilde{S}_i}  \sqrt{r^*}  \right) \right)   ,
	\quad 1\le i \le p,
	\label{eq:thm1errbnd}
	\end{align}
	with probability at least $1-c_1 p^2 T  \exp(- c_2 M^{-1} T^{1/5}) $, where $\phi_{\widetilde{S}_i} = 
\max \limits_{m  \in \widetilde{S}_i  } \sum \limits_{m' \ne m  \in \widetilde{S}_i} w_{m,m'}$, and $c_1, c_2 >0$ depend on the model parameters and the transition kernel.
\end{theorem}
% 
% \sw{
% \begin{remark}
% The term $2\phi_{\widetilde{S}_i}  \sqrt{r^*}$ comes from the error induced by the fusion penalty in the joint estimation. 
% In an ideal case, when the network similarity are known, we can reparameterize the model so that it includes the same parameters that stay the same for all experiments and the experiment-specific parameters that vary across experiments. This greatly reduces the number of parameters involved, particularly when the networks are sparse and similar. Under such reparameterized model, the fusion penalty is no longer needed in the estimation step, which may thus lead to a tighter estimation error bound. However, knowing the true network skeletons is almost impossible in practice. Therefore, in this work, we instead focus on developing the proposed estimation procedure under the unknown network-skeleton case.
% \end{remark}
% }
%
The error bound in Theorem~\ref{theorem1} involves the overall network sparsity $d^*$ and the dissimilarity index $r^*$, suggesting a low prediction error bound for  sparse networks that are similar between experiments. 
%\as{why?} \sw{the bound involves the network sparsity $d^*$ (the max number of non-zero edge of a network), and the similarity index $r^*$ (the max number of dissimilar edges (that are pointed to a node) between networks)  }
%
The network size, $p$, is allowed to grow much faster than the minimum length of the experiments, as long as $\log p = o(\min T^{1/5}_m)$. The number of experiments, $M$, is also allowed to grow faster than the length of the experiments, as long as $ M= o(T^{1/5})$. This condition is likely met in practice, as the total number of experiment is usually not too large and the lengths of experiments are similar; for instance, \citet{Boldingeaat6904} conducted $M=80$ experiments with each experiment consisting of data in 30kHz over 10 seconds. 
The result implies that, compared with methods that separately estimate the network under each experiment, our procedure achieves a faster convergence rate of order $T=\left(\sum_{m=1}^M T_m \right )^{-2/5}$, instead of $\min_{m=1,\dots,M} T^{-2/5}_m $. 
% \sw{
% In an ideal case when the true network skeletons are known and all networks are highly similar with each other, $M$ is allowed to grow in a faster rate than $T$ using a re-parameterized model where the model uses the same parameters that stay the same for all experiments and experiment-specific parameters that vary across experiments.
% }
%\sw{When allowing $M$ grow, it achieves a faster convergence rate in the order of $T=\left(\sum_{m=1}^M T_m \right )^{-2/5}$ instead of  $\min_{m=1,\dots,M} T^{-2/5}_m$. However, in this condition, to achieve the estimation consistency---i.e., the error converges to 0 as $T \rightarrow \infty$, the total sparsity of the $M$ networks needs to be much smaller than $T$---i.e., $\sum_{m=1}^M d^*_m \ll T$, however, this may not be always possible even when all M experiments are equal -- the reason is that when we model the parameters, we specific unique parameter for each condition even the underlying networks are all equal.}
% comment on \phi_{\widetilde{S}}
%
%\as{technically, this only affects the rate if $M$ is allowed to grow}
%\sw{By faster rate, I meant $\left(\sum_{m=1}^M T_m \right )^{-2/5}$ is smaller than $\min_{m=1,\dots,M} T^{-2/5}_m$ as $M>1$. But if $M$ is a pre-specificed number, then $\left(\sum_{m=1}^M T_m \right )^{-2/5}  $ and $\min_{m=1,\dots,M} T^{-2/5}_m$ are in the same order -- is it why I cannot say the rate is faster?}

Theorem~\ref{theorem1} also highlights the effect of using informative weights in the fusion penalty.  
To see this, first note that with normalized similarity weights, %---i.e.,$\sum \limits_{1\le m\ne m' \le M} w_{m,m'}= 1$, 
$$
\phi_{\widetilde{S}_i} = 
\max \limits_{m  \in \widetilde{S}_i  } \sum \limits_{m' \ne m  \in \widetilde{S}_i} w_{m,m'}    \le \max\limits_{1\le m \le M} \sum\limits_{1 \le m' \ne m \le M} w_{m,m'}    \le 1 .
$$ 
Thus, the estimation error is always bounded by $O_p\left(  \frac{1}{\eta }
T^{-2/5}\left( 3\sqrt{d^* }  + 2 \sqrt{r^*}  \right) \right)$, regardless of the choice of weights. 
However, if the weights are correctly specified---i.e., $w_{m,m'}$ are small for pairs of networks that are different---the error bound is improved. Consider, for example, networks corresponding to $M=3$ conditions, where the first two are identical and the third is completely different. Using oracle similarity weights, i.e., $w_{1,2}=1, w_{1,3} = w_{2,3}=0$, we get $\phi_{\widetilde{S}_i} = w_{1,3}+ w_{2,3} =0 $. In contrast, using uninformative weights that treat all networks equally, i.e., $w_{1,2}= w_{1,3}=w_{2,3}=\frac{1}{3}$, we have  $\phi_{\widetilde{S}_i} = w_{1,3}+ w_{2,3} = \frac{2}{3}$.

% \as{this is not very clear and given its importance needs to be clarified; especially since in the intro we say we `show' this}
% \sw{
% When the similarity weights are appropriately specified, $w_{m,m'}$ is small for experiments $m$ and $m'$ whose underlying networks are quite different. Thus, 
% $\phi_{\widetilde{S}_i} = 
%  \max \limits_{m  \in \widetilde{S}_i } \sum \limits_{m' \ne m  \in \widetilde{S}_i} w_{m,m'}$ is small. need to add examples to illustrate the oracle and poor weights case. 
% }
%especially when there exists networks that are quite different---i.e., when $r^*$ is large. 
% see proof theorem 1, \eqref{eq:bound_D} 

\begin{corollary}\label{corollary1}
Assume the setting of Theorem~\ref{theorem1} and in particular Assumptions~\ref{assumption1}--~\ref{assumption4}. Then, the result \eqref{eq:thm1errbnd}  holds for
\begin{enumerate}
\item linear Hawkes processes, with positive background intensities, estimated using the least square loss, $\ell(a,b) = (a-b)^2 $;
\item non-linear Hawkes processes, with exponential-link function, $g(\cdot) = \exp(\cdot)$, estimated using the negative log likelihood loss, $\ell \left ( a,  b \right )  = - a \log (b)  + b $.
\end{enumerate} 
\end{corollary}
%Corollary~\ref{corollary1} holds because Condition~\ref{def:RSC} and~\ref{def:tail_bound} in Theorem~\ref{theorem1} are satisfied for each case of the link function and loss function pair. 

To establish the edge selection consistency, we next introduce an additional assumption. 
\begin{assumption}
	\label{assumption5}
	There exists $\tau >0 $ such that for $1\le i \le p$
	\begin{align*}
	\min_{  \beta^{(m)}_{ij} \in  \bigcup_{m=1}^M S_i^{(m)}  } \beta^{(m)}_{ij} \ge \beta_{min} > 2\tau. 
	\end{align*}
\end{assumption}
Assumption~\ref{assumption5} is known as the `$\beta$-$\min$ condition' \citep{buhlmann2011} and requires sufficient signal strength for the true edges in order to distinguish them from $0$. 
%\as{we have not really defined the thresholded estimator, or did I miss it?}
%
To infer the connectivity patterns, we consider the \textit{thresholded connectivity estimator} 
$$ 
\widetilde{\beta} ^{(m)}_{ij}  = \widehat{\beta} ^{(m)}_{ij} \mathbf{1} \left( \left |\widehat{\beta} ^{(m)}_{ij} \right | >  \tau \right ),  \quad 1\le i,j \le p .$$ 
Thresholded estimators are particularly appealing for high-dimensional network estimation \citep{shojaie2012}, as they offer consistent variable selection under mild assumptions \citep{vandeGeer2011EJS}. Denoting the estimated and true edge set by $\widehat{S}^{(m)} =\left  \{ (i,j): \widetilde{\beta}^{(m)}_{ij}  \ne 0 , 1\le i,j \le p \right \}$ and 
$S^{(m)} =\left  \{ (i,j): \beta^{(m)}_{ij}  \ne 0 , 1\le i,j \le p \right \}$, respectively, we next establish the consistency of the estimated edge set.
\begin{theorem}
	\label{theorem2}
Under the same conditions in Theorem~\ref{theorem1}, suppose Assumption~\ref{assumption5} is also satisfied with
$\tau = O\left(  \frac{1}{\eta }
	T^{-2/5}\left( 3\sqrt{d^*} + 2\max \limits_{1\le i \le p} \phi_{\widetilde{S}_i}    \sqrt{r^*}  \right) \right)$. 
Then,  
	\begin{align*}
	\mathbb{P}\left ( \bigcap_{m=1}^M \left \{ \widehat{S}^{(m)} = S^{(m)} \right \}  \right )
	\ge  1-c_1 p^2  T \exp(- c_2 M^{-1}T^{1/5})  ,
	\end{align*}	
 where $\phi_{\widetilde{S}_i} = 
	\max \limits_{m  \in \widetilde{S}_i} \sum \limits_{m' \ne m  \in \widetilde{S}_i} w_{m,m'}$, and $c_1, c_2 >0$ depending on the model parameters and the transition kernel. 
\end{theorem}

% We add the proofs of both theorems and the corollary in Appendix~\ref{sec:proofs}.

%%%%%%%%%%%%%
\section{Multi-Experiment Inference}\label{sec:test}
%%%%%%%%%%%%%
%[general idea]
%Evaluating the uncertainty of the network estimates is critical in scientific applications. 
%Statistical inference procedure has recently been developed for high-dimensional Hawkes processes \citep{wang2020statistical}. However, implementing the method directly on each of the networks at all experiments involves a large number of tests. As a result, the testing power can be low, especially when the number of experiments is large. 
%Moreover, the tests associated with the edges under the same experiment are complexly dependent because the same data are used to construct the tests.
%
In this section, we develop a hierarchical testing procedure for edges of all $M$ networks, using the hierarchy learned from the network similarity weights in Section~\ref{sec:estimation}. Taking advantage of the hierarchical structure, our procedure greatly reduces the number of tests, resulting in improved power while controlling the family-wise error rate (FWER). Importantly, the FWER control is achieved under arbitrary hierarchical structure determining dependencies among tests. 

In the following, we first discuss how the hierarchy is learned from the similarity weights, and then present our testing procedure. Results from this section are proved in Appendix~\ref{sec:proofs}.

% \as{I think this section needs a bit of restructuring:\\
% 1. I would make the procedure a 2-step procedure only consisting of steps 2 and 3 here. This is because (a) the first step below is not really part of the inference procedure and (b) the last step is not really a step -- you can say that we repeat the following two steps for all $p^2$ edges.\\
% 2. We need to formally define the step 1 below based on clustering -- see eg my paper with Kean Ming and Daniela where we show how thresholding the covariance matrix is equivalent to single linkage hierarchical clustering.\\
% 3. I think it would be helpful to add a picture to illustrate the inference procedure over the dentrogram -- again see the paper mentioend above\\
% 4. Finally, you may want to put the procedure in an algorithm environment so you can refer to it later in the theorems, but I leave it to you to decide on this one}

%\sw{below draft to build binary tree from bottom-up for the the hierarchy. } 

\subsection{Hierarchy for Multi-Experiment Inference}
Our hierarchical testing procedure for $M$ experiments utilizes a binary tree, in which the left child node is always a leaf node corresponding to a single hypothesis. 
This binary tree has $M$ levels, where the level of each node equals one plus the number of connections between the node and the root of the tree. For example, the root is at level $1$ and the two nodes at the bottom of the tree are at level $M$. 
To build our tree from the network similarity weights, we use a procedure similar to hierarchical clustering with single linkage, but with minor differences to facilitate hierarchical inference. 

We build the tree from bottom up. The two nodes at the bottom level (level $M$) correspond to a single experiment each. We assign to the \textit{right} node the experiment whose network has the fewest edges. Specifically, denoting this experiment as $t$, the right node at bottom is assigned the index set, $\mathcal{L}_{R,M} =\{ t \}$. We then assign to the \textit{left} node the experiment whose network is most similar to that of experiment $t$. In other words, let
\[
    s = \argmax_{ 
    s \in \mathcal{L}_1 \backslash \mathcal{L}_{R,M} }
    d( s, t ), 
\]
where $\mathcal{L}_1=\{ 1,\dots, M\}$ and $d(s,t)$ is the similarity between two experiments according to the network similarity weights. 
The left node at bottom is assigned the index set, $\mathcal{L}_{L,M} =\{ s \}$.
Next, we merge the index sets of nodes at level $M$ and assign it to the right node at the upper level; that is, $\mathcal{L}_{R,M-1} = \mathcal{L}_{R,M} \bigcup \mathcal{L}_{L,M} = \{ s, t\}$. The left node at level $M-1$ will be assigned an individual experiment whose network is the closest to $ \mathcal{L}_{R,M-1}$ based on single linkage distance (in case of ties, one experiment is chosen at random). Formally, at each level $l = M-1, \ldots, 2$, the single experiment in the left node is given by
% \as{can you remind me why this is different than HC with single linkage}
% \sw{we start with the network of the fewest edges as the right node at the bottom, and the dendrogram is always a binary tree; but for the single-linkage HC, the algorithm does not necessarily start with the network with the fewest edge and the resulted dendrogram is not always a binary tree---e.g., when more than one networks are equally closest to the existed cluster on the right node of the tree.}
\begin{align} \label{eq:find_left_node}
      s = \argmax_{ 
    s \in \mathcal{L}_1 \backslash \mathcal{L}_{R,l}  
    }
    \max \limits_{ 
    t \in \mathcal{L}_{R,l} }
    d( s, t ). 
\end{align}
%
% Start from the bottom nodes, we assign a set of experiments to each node of the tree. Specifically, the right node at bottom (level $M$) is assigned to index set $\mathcal{L}_{R,M} =\{ t \}$, where $t$ corresponds to the experiment whose network has the fewest edges. The left node is assigned to the index set $\mathcal{L}_{L,M} =\{ s \}$, where $s$ corresponds to the experiment whose network is most similar to that of experiment $t$; i.e, 
% \begin{align} \label{eq:find_left_node}
%       s = \argmax_{ 
%     s \in \mathcal{L}_1 \backslash \mathcal{L}_{R,M}  
%     }
%     \max \limits_{ 
%     t \in \mathcal{L}_{R,M} }
%     d( s, t ), 
% \end{align}
% where $\mathcal{L}_1=\{ 1,\dots, M\}$ and $d(s,t)$ is the similarity between two experiments according to the network similarity weights.
% We then merge the index sets of both nodes into one set and assigned it to the right child node at the upper level---i.e., $\mathcal{L}_{R,M-1} = \mathcal{L}_{R,M} \bigcup \mathcal{L}_{L,M} = \{ s, t\}$. Next, we assign the left node at level $M-1$ the index set with a single element corresponding to the experiment that is closest to $ \mathcal{L}_{R,M-1}$ as defined in \eqref{eq:find_left_node}. 
This procedure is repeated until we reach the root of the binary tree, which is assigned the total index set $ \mathcal{L}_1$. 

% The above procedure can incorporate different similarity measures. Throughout this paper, we consider the network similarity measures discussed in Section~\ref{sec:estimation}. In particular, we refer to resulting binary tree $\mathcal{D}$ as either the \textit{oracle} or \textit{empirical} if it is built using the oracle network similarity, $d(s,t) = d^o_{s,t}$ in \eqref{def:oracle_dist}, or the empirical network similarity, $d(s,t) =d^e_{s,t}$ in \eqref{eq:empirical_similarity}. We illustrate an example of the binary tree construction based on a similarity matrix over 4 conditions in Figure~\ref{fig:BT}. 

%\as{Good idea. I modified the previous paragraph as follows. Let me know what you think.} \sw{the follows looks great}

\begin{figure}[t]
\centering
\begin{minipage}[b]{.35\textwidth}
\includegraphics[width=.95\textwidth]{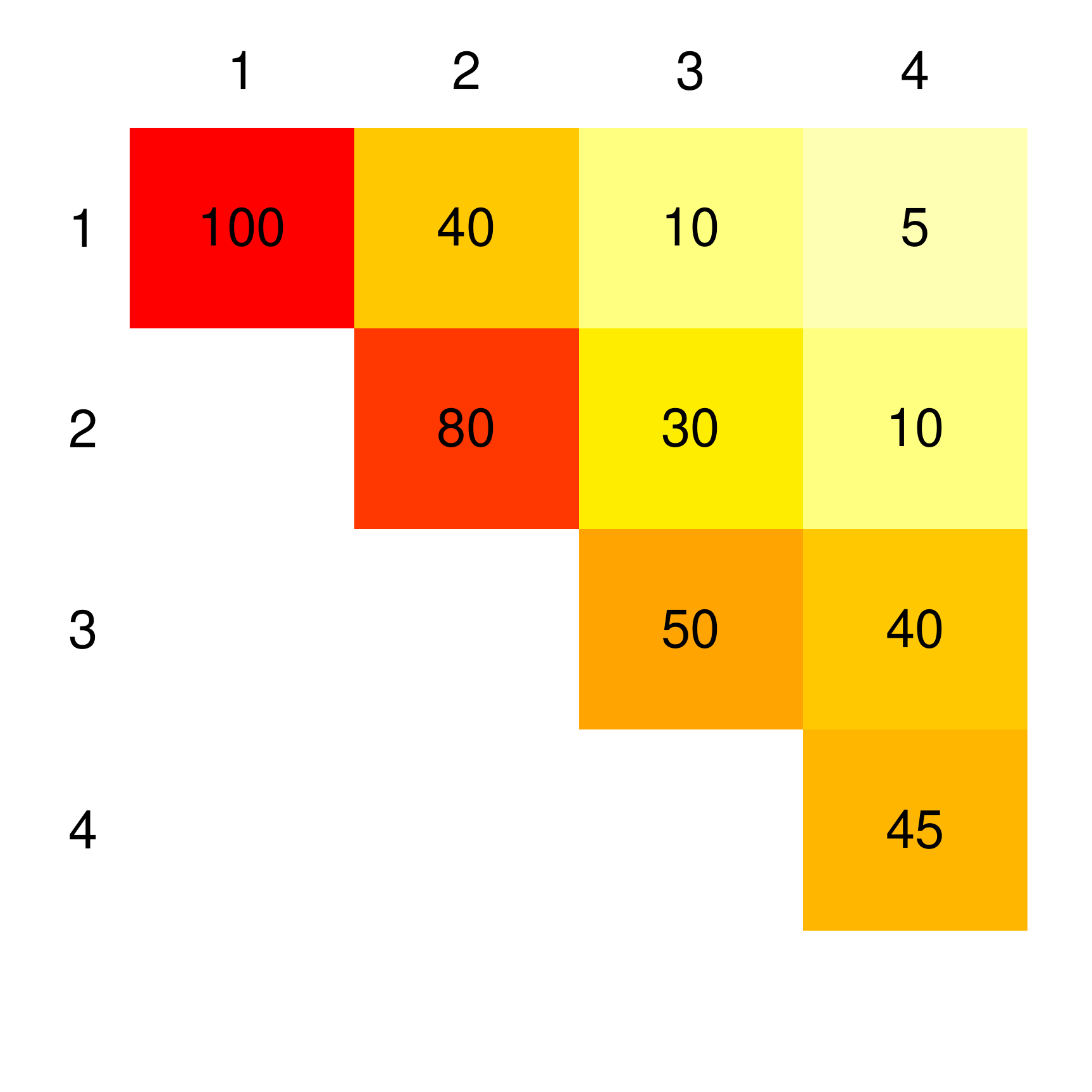}
\caption*{Similarity matrix} 
\end{minipage}\hfill
\begin{minipage}[b]{.65\textwidth}
\includegraphics[width=0.95\textwidth]{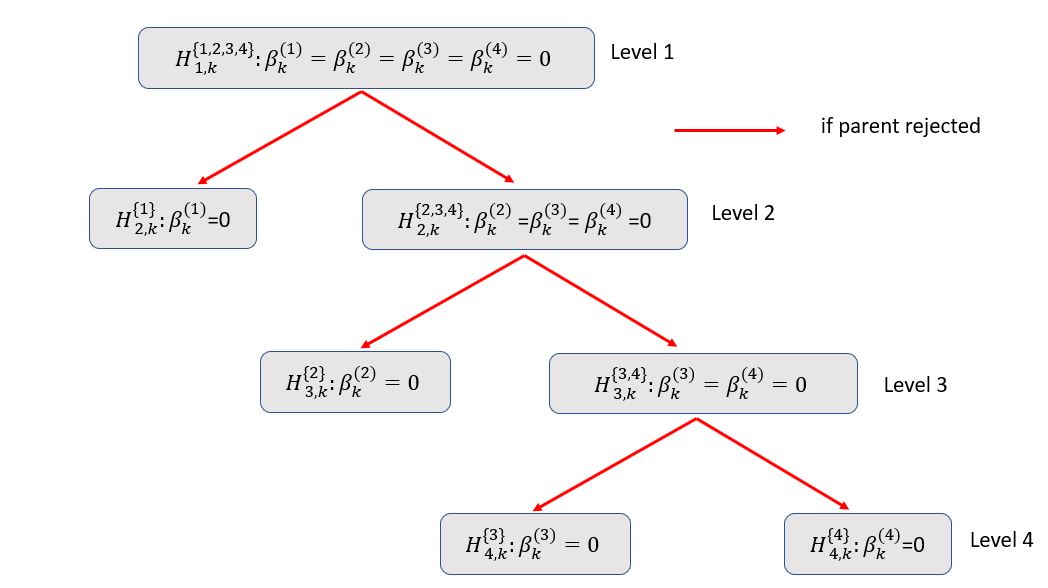}
\caption*{Illustration on hierarchical testing procedure}
\end{minipage}
    \caption{
    The similarity matrix corresponds to the number of common edges of networks under 4 conditions. Specifically, 40 edges are shared between Conditions 3 and 4, and 4 edges are shared between Conditions 1 and 4. Condition 4 has the fewest edges (45) among all conditions. A $4$-level binary tree (right) is then built according to the network similarity to guide the hierarchical testing procedure.}
    \label{fig:BT}
\end{figure}

The above procedure can incorporate different similarity measures, including those discussed in Section~\ref{sec:estimation}. Of particular interest is the binary tree built using the empirical network  similarity, $d(s,t) =d^e_{s,t}$ in \eqref{eq:empirical_similarity} which is based on the number of shared edges between networks. 
We refer to this tree as the \textit{empirical tree}. An example based on a similarity matrix over 4 conditions is given in Figure~\ref{fig:BT}.

\subsection{Hierarchical Inference}
Given the binary tree $\mathcal{D}$ from the previous section, we next describe the hierarchical testing procedure.
%and show that this procedure is more powerful in testing the edge coefficients over all experiments while controlling FWER. 
%[hierarchical testing description]
For ease of notations, we index the $p^2$ edge coefficients of a $p$-variate network at condition $m$ as $ \beta^{(m)}_{1},\dots,\beta^{(m)}_{p^2}$, for $m=1,\dots, M$. Besides the binary tree, $\mathcal{D}$, the algorithm takes in the critical values at each level of the tree, $\{ \alpha_l\}_{l=1}^M$. 

Our procedure, summarized in Algorithm~\ref{alg:HT}, is applied separately to each coefficient $\beta_k, k \in\{1,\dots, p^2 \}$. At each node of the hierarchy $\mathcal{D}$, we test the global hypothesis that all the edge coefficients corresponding to the experiments indexed by the node are 0; that is, we test $H_{l,k}^{ \mathcal{L} } : \beta_k^{(m)} =0 , m \in \mathcal{L}$ where $ \mathcal{L} \in \{\mathcal{L}_1, \mathcal{L}_{L, l}, \mathcal{L}_{R, l}\}$, depending on whether the node is the root of the binary tree, or the left or right child node at level $l$, respectively. Our procedure starts by testing the hypothesis at the root of the tree. If $H_{l,k}^{ \mathcal{L}_1 }$ is rejected, we move down to the next level of the tree and separately test the hypotheses assigned to each of the child nodes. The process continues until we reach a level $m \in \{1,\ldots,M\}$ such that $H_{m,k}^{ \mathcal{L}_{R,m} }$ is not rejected. %This \textit{hierarchical testing} procedure is summarized in Algorithm~\ref{alg:HT}. 
%which requires $p^2$ to $p^2 (2M-1)$ operations. 

Consider, for example, testing the edge coefficients for the $M=4$ networks corresponding to the hierarchy defined by the binary tree in Figure~\ref{fig:BT}. 
Let $Z$ be the $p^2 \times M$ matrix of rejection indicators. 
%\as{do we use $I$ for identify matrix anywhere? if yes, we'd need a different notation} \sw{maybe use $Z$?}
For each edge coefficient $\beta_k, k \in\{1,\dots, p^2 \}$, we start from the root of the tree and test $H^{ \{ 1,2,3,4 \} }_{1,k}: \beta_k^{(1)} =  \beta_k^{(2)} = \beta_k^{(3)} =\beta_k^{(4)} = 0$. If $H^{ \{ 1,2,3,4 \} }_{1,k}$ is rejected, then we move down to the next level and separately test 
$H^{ \{ 1 \} }_{2,k}: \beta_k^{(1)} = 0 $ and $H^{ \{ 2,3,4 \} }_{2,k}: \beta_k^{(2)} = \beta_k^{(3)} =\beta_k^{(4)} = 0 $. 
If we reject $H^{ \{ 1 \} }_{2,k} $ then $Z_{k, 1} = 1$; otherwise, $Z_{k, 1} = 0$. We continue this process on the right branch by testing $H^{ \{ 2,3,4 \} }_{2,k} $. 
% If we reject $H^{ \{ 2,3,4 \} }_{2,k} $ then we move down to test $H^{ \{ 2  \} }_{3,k} $ and $H^{ \{ 3,4 \} }_{3,k} $;  otherwise, $I_{k, 2} =I_{k, 3}= I_{k, 4}= 0$. 

\begin{algorithm}[t] 
% \as{The notation for the algorithm is too busy. Can we simplify it?} \sw{maybe use notation $H_{l,k}^{\mathcal{L}_{L,l}}$ instead of its detailed version $H_{l,k}^{\mathcal{L}_{L,l}}: \beta^{(m)}_i = 0, m \in \mathcal{L}_{L,l}$ and let audience find the definition in the text? }
    \caption{Hierarchical Testing Procedure for Multi-Experiment Networks}
	\begin{algorithmic} \label{alg:HT}
	\STATE{\textbf{input}:   $\mathcal{D}$, $\{ \alpha_l\}_{l=1}^M$;}
	\STATE{\textbf{initialization}: 
	             rejection matrix $Z$ ;
	             %\sw{ $ Z = \{ Z_{k,m}=0 \}_{1\le k \le p^2, 1\le m \le M} $ };
	             }
	\FOR {$k=1,\dots, p^2$}
% 	\STATE{
% 	\textbf{root}: calculate $p$-value, $P_k^{\mathcal{L}_1}$, on $H_{1,k}^{\mathcal{L}_1}: \beta^{(m)}_k = 0, m\in \mathcal{L}_1 = \{1,\dots, M \} $; 
% 	}
	\STATE{
	\textbf{root}: calculate the $p$-value  $P_k^{\mathcal{L}_1}$ for  $H_{1,k}^{\mathcal{L}_1}$; 
	}	
	\IF {$P_k^{\mathcal{L}_1} \le \alpha_1$ and $M > 1$ } 
    	\FOR {$l=2,\dots, M$}
    % 	\STATE{
    % 	\textbf{left node}: calculate $p$-value, $P_k^{\mathcal{L}_{L,l}}$, on $H_{l,k}^{\mathcal{L}_{L,l}}: \beta^{(m)}_i = 0, m \in \mathcal{L}_{L,l}$; if 
    % 	$ P_k^{\mathcal{L}_{L,l} } \le \alpha_l $, 
    % 	set $Z_{k,m} = 1$, for $m \in  \mathcal{L}_{L,l}$ ; 
    % 	}
    	\STATE{
    	\textbf{left node}: calculate the $p$-value $P_k^{\mathcal{L}_{L,l}}$ for $H_{l,k}^{\mathcal{L}_{L,l}}$; if 
    	$ P_k^{\mathcal{L}_{L,l} } \le \alpha_l $, 
    	set $Z_{k,m} = 1$, for $m \in  \mathcal{L}_{L,l}$ ; 
    	}    	
    % 	\STATE{
    % 	\textbf{right node}:
    % 	calculate $p$-value, $P_k^{\mathcal{L}_{R,l}}$, on $H_{m,k}^{\mathcal{L}_{R,l}}: \beta^{(m)}_i = 0, m \in \mathcal{L}_{R,l}$ ; 
    % 	stop the loop over $l$ if $P_k^{\mathcal{L}_{R,l}} > \alpha_l $; otherwise, set $Z_{k,m} = 1$ for $m\in \mathcal{L}_{R,l} $ if $l=M$;
    % 	}
    	\STATE{
    	\textbf{right node}:
    	calculate the $p$-value  $P_k^{\mathcal{L}_{R,l}}$ for $H_{m,k}^{\mathcal{L}_{R,l}}$ ; 
    	if $P_k^{\mathcal{L}_{R,l}} > \alpha_l $, stop the loop at level $l$; otherwise, if $l=M$, set $Z_{k,m} = 1$ for $m\in \mathcal{L}_{R,l} $;
    	}
    	\ENDFOR
    \ENDIF
	\ENDFOR
	\RETURN $Z$
	\end{algorithmic}
\end{algorithm}

Algorithm~\ref{alg:HT} can accommodate $p$-values  from any valid test of edge coefficients. Here, we use the \textit{de-correlated score statistics} for testing $\beta^{(m)}_{ij}=0$ \citep{wang2020statistical}, defined as 
% Specifically, the method calculates the \textit{de-correlated score} statistics
\begin{align*}
S^{(m)}_{ij} &= \frac{1}{T_m} \int_0^{T_m} \epsilon^{(m)}_i(t) \widetilde{x}_j^{(m)}(t) dt ,  
\end{align*}
where $ \epsilon^{(m)}_i(t) = \frac{dN^{(m)}_i(t)}{dt} - \lambda^{(m)}_i(t) $, and $\widetilde{x}_j^{(m)} $ is the de-correlated column, obtained from $x_j^{(m)}(t) $ after removing its projection on the other columns, $x^{(m)}_{-j}(t)$. 
Denoting $ \Upsilon^{(m)}_j  = \frac{1}{T} \int_0^T \left (\widetilde{x}_j^{(m)}(t)  \right )^2 dt $, and $V^{(m)}_{ij}  = \sqrt{T} \left( \Upsilon^{(m)}_j  \right)^{-1/2} S^{(m)}_{ij}  $, $ V^{(m)}_{ij} \rightarrow_d \mathcal{N}(0,1) $. 
% It has been shown that $ V^{(m)}_{ij} \rightarrow_d \mathcal{N}(0,1) $, where $V^{(m)}_{ij}  = \sqrt{T} \left( \Upsilon^{(m)}_j  \right)^{-1/2} S^{(m)}_{ij}  $ and $ \Upsilon^{(m)}_j  = \frac{1}{T} \int_0^T \left (\widetilde{x}_j^{(m)}(t)  \right )^2 dt $.  
% Since $V^{(m)}_{ij}$ involves unknown model parameters in its calculation, in practice, consistent estimates on those parameters (e.g., via penalized regression) can be used instead. We refer the audience for the original paper for details.
Thus, the global hypothesis at the root,  $H_{1,k}^{\mathcal{L}_1}$, can be tested using the test statistic 
\begin{align*}
   U^{\mathrm{total}}_{ij} = \sum \limits_{  m \in \mathcal{L}_1 } \left (V^{(m)}_{ij} \right )^2 \rightarrow_d \chi^2_M ,
\end{align*}
and the corresponding $p$-value is approximately $\mathbb{P} \left(   \chi^2_M \ge U^{\mathrm{total}}_{ij}  \right) $.
By construction, such a test is powerful when many $\beta_{ij}^{(m)} \ne 0, m \in \mathcal{L}_1$. 
When, on the contrary, a small number of edge coefficients are nonzero, 
%Notice that to make such test valid, $M$ needs to be a pre-specified value. 
%
an alternative and more powerful test can be  constructed based on the maximum of $\left (V^{(m)}_{ij} \right )^2$; that is, 
\begin{align*}
  U^{\max}_{ij} = \max_{  m \in \mathcal{L}_1 } \left (V^{(m)}_{ij} \right )^2 .
\end{align*}
It follows from the asymptotic distribution of $V^{(m)}_{ij}$ \citep[e.g.,][pp~156]{Embrechts1997} that as $M \rightarrow \infty$, 
% \begin{align*}
%     a_M(  U_{ij} - b_M ) &\rightarrow_d G ,  \\
%     a_M = 1/2   ,  
%     \quad \as{b_n} &= 2( \ln M + (d-1)\ln\ln M - \ln \Gamma(d) ), 
% \end{align*}
\begin{equation*}
    a_M(  U^{\max}_{ij} - b_M ) \rightarrow_d G , 
\end{equation*}
where $a_M = 1/2$, $b_M = 2( \ln M + (d-1)\ln\ln M - \ln \Gamma(d) )$ 
%\as{should it be $b_M$?} 
and $G$ follows Gumble distribution.
%---$\mathbb{P}( G \le x) = \exp( - \exp(-x)), x \in \mathbb{R}$.
% By construction, $U^{\max}$ requires a large $M$ to obtain valid $p$-values calculated from the limiting distribution---i.e., the Gumble distribution. 

{
We next introduce an ideal binary tree for inference, which we refer to the  \textit{oracle} tree. In addition to being built based on the oracle network similarity distance, the main difference between this  tree and the empirical tree introduced earlier is that the oracle tree is edge-specific. Specifically, for each of the $p^2$ edge coefficients to be tested, the binary tree is built using the similarity distance $d^{(k)}(s,t) = |\beta^{(s)}_k - \beta^{(t)}_k |$. Thus, zero coefficients are always places at the bottom right of the oracle tree. Clearly, this information (and the oracle tree) is hardly available in practice, and is primarily used as a theoretical device in the next result to establish the control of the family-wise error rate (FWER) under arbitrary dependencies between tests.
}

% \sw{
% Next, we introduce the \textit{oracle} binary tree, which is edge-specific. Specifically, for each of the $p^2$ edge coefficients to be tested, the binary tree is built using the similarity distance $d^{(k)}(s,t) = |\beta^{(s)}_k - \beta^{(s)}_k |$. Consequently, the oracle binary tree always puts the zero-coefficients to its bottom right. Although such oracle tree is hardly available in practice, it allows our hierarchical procedure controls the family-wise error rate (FWER) under arbitrary dependencies between tests (as shown below).} 

\begin{theorem}\label{fwer_oracle}
%With known similarity of the multi-experiment networks \as{what does this mean? formalize}, the hierarchical testing procedure \as{give more specific ref to it, eg defined in...} 
The hierarchical testing procedure in Algorithm~\ref{alg:HT} with the oracle binary tree controls the FWER for testing all $p^2M$ hypotheses at level $\alpha$ when using the critical value 
\begin{align*}
	\alpha_l = \frac{\alpha}{p^2}\frac{M-l+1}{M}, \quad l = 1,\dots, M . 
\end{align*}
\end{theorem}

% \begin{remark}
% Our proposed procedure can be applied to any subset, $\mathcal{J}$, of the $p^2$ edges, where in Theorem~\ref{fwer_oracle} we take $\alpha_m = \frac{\alpha}{|\mathcal{J}|}\frac{M-l+1}{M}$. 
% \end{remark}

For a single experiment, the $\alpha_l$ in Theorem~\ref{fwer_oracle} amounts to the usual Bonferroni correction. However, in multi-experiment settings, our procedure uses a less stringent critical value than that Bonferroni correction, as $\frac{\alpha}{p^2}\frac{M-l+1}{M} <  \frac{\alpha}{M p^2}$ for $l < M$. This makes the procedure more powerful in practice, particularly for sparse networks when most tests are carried out at shallow levels of the tree. 
Unlike existing hierarchical testing methods \citep[e.g.][]{Yekutieli2008,lynch2016procedures} that control the error rate all the tests involved, our procedure controls the error among the hypothesis associated with the leaves of the tree---this is exactly the set of hypotheses of  interest in our multi-experiment network inference problem. Lastly, the procedure can also be applied to any subset, $\mathcal{J}$, of the $p^2$ edges by taking $\alpha_l = \frac{\alpha}{|\mathcal{J}|}\frac{M-l+1}{M}$ in Theorem~\ref{fwer_oracle}. 

Theorem~\ref{fwer_oracle} assumes an oracle binary tree, which is unavailable in practice. In such cases, the data-driven similarity in Section~\ref{sec:estimation} can be used to create an empirical binary tree. 
The next result shows that, for large and sparse networks, our procedure is robust to potential misspecification of the binary tree.
\begin{theorem}\label{fwer_general}
%Then, with \as{unknown similarity of the multi-experiment networks AGAIN, WHAT DOES THIS MEAN?}, the \textit{hierarchical testing} procedure
The testing procedure in Algorithm~\ref{alg:HT} with a binary tree built based on arbitrary network similarity controls FWER at level $  \alpha \left( 1+ \frac{d^* M (M-1) }{2 p}  \right) $ when using critical values
	\begin{align*}
	\alpha_l = \frac{\alpha}{p^2}\frac{M-l+1}{M}, \quad l =  1,\dots, M .
	\end{align*}
\end{theorem}
Theorem~\ref{fwer_general} implies that when $d^* M(M-1) = o(p)$, FWER is controlled at $  \alpha \left( 1+ o(1)  \right) $ regardless of the hierarchy used in the testing procedure. This condition is met when the underlying network is sparse, i.e, $d^*= o(p)$, and the number of experiment is not too large, i.e., $M = o(\sqrt{p})$. Our proof in Appendix~\ref{sec:proofs} indicates that the unique construction of the binary tree, where the left child node is always a leaf, is critical for achieving this robustness. 

\section{Simulations}\label{sec:sims}
%%%%%%%%%%%%%

% \begin{figure}[t]
% 	\centering
% 	\includegraphics[width=\linewidth, clip=TRUE, trim=0mm 10mm 0mm 0mm]{network_star_circle.pdf}
% 	\caption{Networks of $p=100$ processes under $M=3$ experiments. Network~1 (left) consists of 20 circles, Network~3 (right) consists of 20 stars, and Network~2 (middle) is a mix of 18 circles and 2 stars.\sw{remove this plot to save space?} \as{yes}} 
% 	\label{fig:network_star_circle}
% \end{figure}

%%%
\begin{figure}[t!]
\subfloat[]{\includegraphics[width=0.5\linewidth]{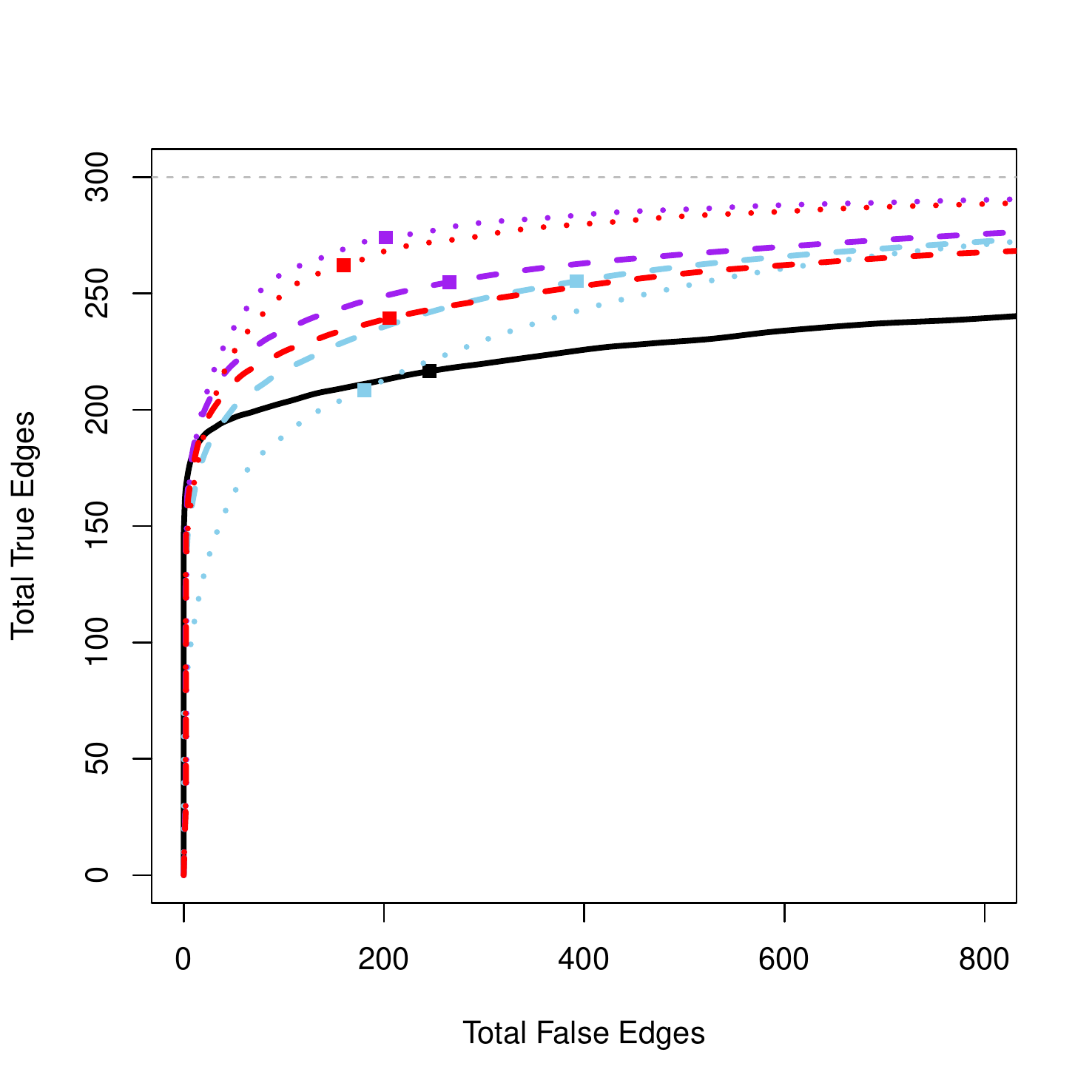}} 
\subfloat[]{\includegraphics[width=0.5\linewidth]{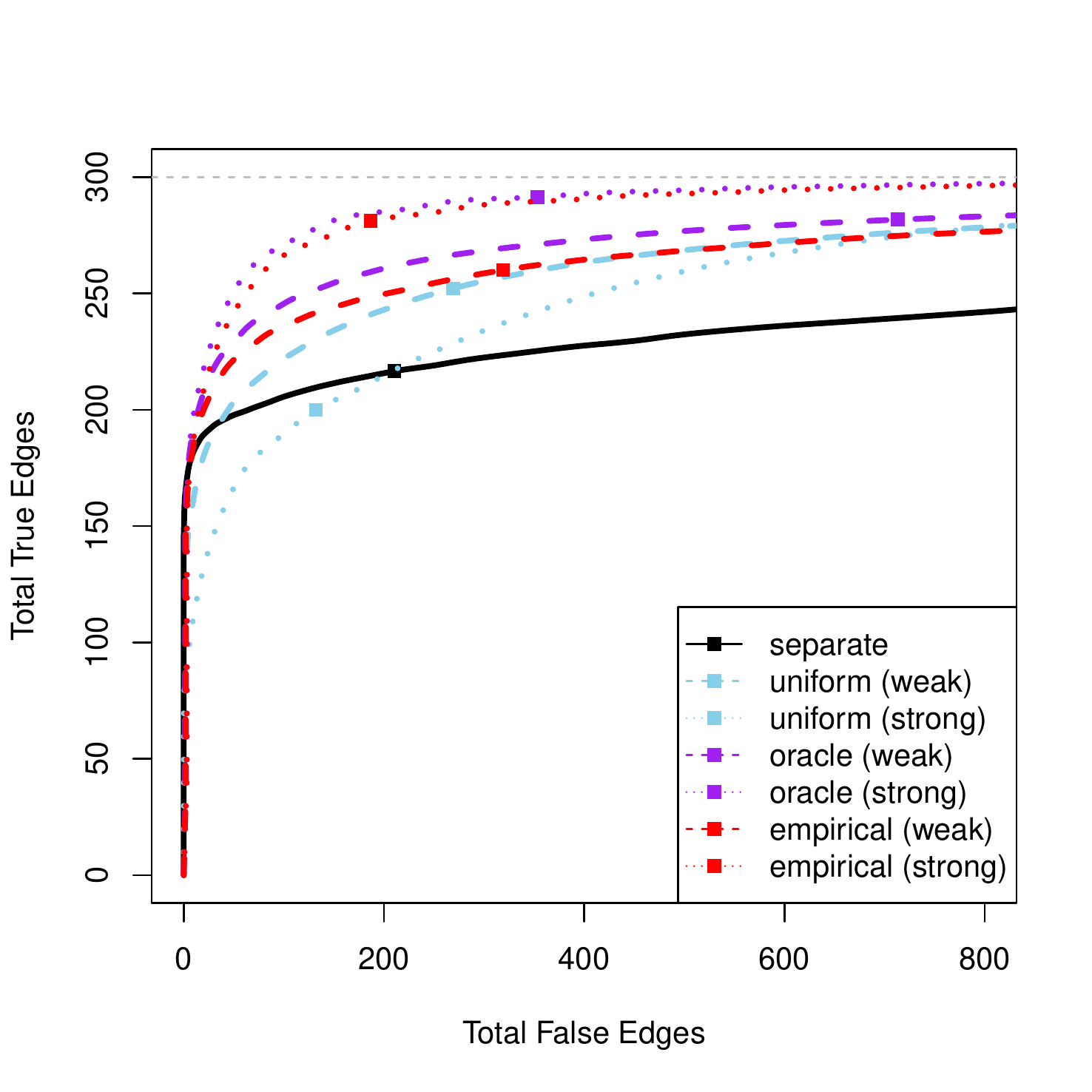}}
	\caption{Edge selection performance of the proposed joint estimation method in a simulation study focused on inferring edges in 3 networks of linear Hawkes processes. The plots show average number of true positive and false positive edges, over 100 simulation runs, for the joint estimation method with different choices of weights, compared to separate estimation of each network. Weight strategies include oracle, empirical and uniform weights. Solid squares ($\blacksquare$) correspond the choice of tuning parameter using eBIC. (a): Network~2 shares 90\% edges with Network 1 and 10\% with Network 3 as in Figure~\ref{fig:network_star_circle}. (b): Network 2 is the same as Network~1. 
	%\as{These colors look better, but please change the green to orange and use the same color in other plots too}
	}
		\label{fig:lstpfp2}
\end{figure}

\subsection{Performance of Joint Estimation}\label{sec:sims-est}
We first investigate the edge selection performance of the proposed joint estimation procedure. 
We consider $M=3$ networks of $p=100$ linear Hawkes processes. The networks are designed such that Networks~1 and 2 are much more similar to each other than Network~3. Specifically, Network~1 and 3 consists of 20 5-node  circles and stars, respectively, and Network~2 is a mix of 18 circles and 2 stars (see Figure~\ref{fig:network_star_circle} in Appendix~\ref{sec:addition}).  
The edge coefficients of circles and stars are set to be 0.3 and 0.6, respectively. 
The background intensity is set to 0.2 for all nodes in all experiments. 
The transfer kernel function is chosen to be $\exp(-t)$, for all nodes in all experiments. This setting satisfies our assumptions of a stationary Hawkes process under each experiment. The time periods, $T_m$, are 200, 500, 300 for $m=1,2,3$, respectively.  

We consider three weight choices: informative weights based on the true networks (oracle) and the cross-correlation method of Section~\ref{sec:estimation} (empirical), and uniform weights that treat all networks equally. 
We consider weak and strong fusion penalties---$\rho_2=\rho_1$ and $\rho_2 = 10\rho_1$---and compare them to separate network estimation. %--i.e. $\rho_2=\rho_1$ and $\rho_2 = 10\rho_1$. 
%In addition we compare our proposed method with the method that separately estimates the network under each experiment. %--i.e. $\rho_2 =0$. 

Simulation results are summarized in Figure~\ref{fig:lstpfp2}. 
It can be seen that our proposed empirical weights perform very similar to the oracle weights and both versions of informative weights greatly improve the edge selection performance compared with the uniform weights. 
% \sw{Moreover, while the benefits are more pronounced with larger fusion penalty, the improvements from informative weights are robust to the choice of tuning parameter for the fusion penalty.} \as{is that really true?} \sw{I wanted to say with the informative weights, giving too large tuning parameter will not impact the performance; while using the uniform weights, too large tuning parameter will harm the performance -- I think I can remove this statement since later I mentioned this when talking about the uniform weights.}
Moreover, while the advantages of the informative weights are clear, even uniform weights can perform better than method that estimates each network separately; however, with uniform weights, the performance of the method is sensitive to the choice of the tuning parameter for the fusion penalty. %Identifying the optimal tuning parameter in this case may be challenging. 
The benefit of our estimation procedure depends on the similarity between networks. For example, when we alter Network~2 to be the same as Network~1, we observe greater advantages of our method compared to estimating each network separately or using the uniform weights (Figure~\ref{fig:lstpfp2}(b)). 
Additional simulation results in Appendix~\ref{sec:addition} (Figure~\ref{fig:lstpfp3}) indicate that the performance of our joint estimation procedure improves with increasing number of experiments,  if the additional experiments are similar to some of the existed ones.  

\subsection{Performance of Hierarchical Inference}
\begin{figure}[!t]
	\centering
	\includegraphics[width=1\linewidth]{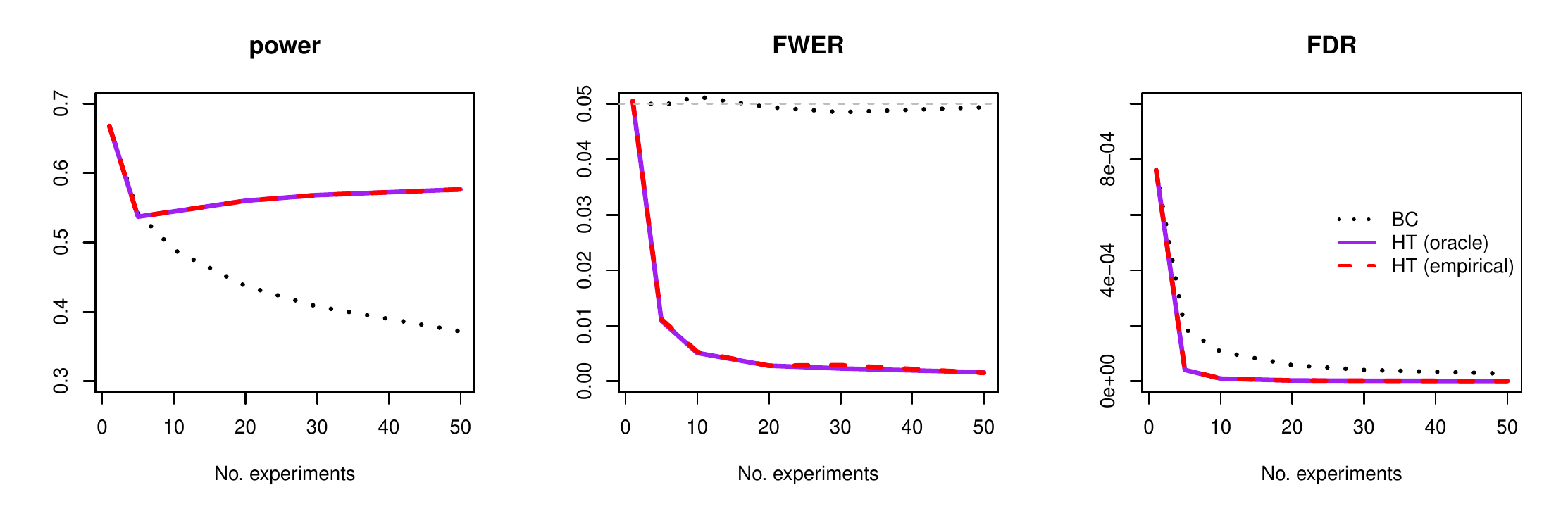}
	\caption{Power, FWER and false discovery rate (FDR) for Bonferroni correction and the proposed hierarchical testing procedure using the oracle and empirical binary trees over 1000 simulation runs. The FWER is controlled at $\alpha =0.05$ (gray dash line). % \as{pls match the colors with the previous plot (dashed red for empirical, solid purple for oracle and dotted black for BC}
%	\as{Let's use red and orange as in the previous figure (matching the colors to be the same, red for empirical and orange for oracle) but let's also use some symbols on the lines with jitter to distinguish the lines a bit.} \sw{chat to see the best symbols for lines? I tried add 'x' on points of the oracle ones, but it is still not that different... }
}
	\label{fig:HT_plot}
\end{figure}

Next, to evaluate the performance of the hierarchical testing procedure, we consider $M=1,5,10, 20, 30, 50$ experiments where the first $M-1$ networks are the same as Network~1 in the previous subsection and the $M$th network is the same as Network~3. We compare our proposed procedure with Bonferroni correction in terms of power, control of FWER, and false discovery rate (FDR). We run our procedure using the oracle and empirical binary trees. As in Figure~\ref{fig:lstpfp2}, we observe similar performances using both types of weights. The results in Figure~\ref{fig:HT_plot} indicate that our hierarchical inference procedure controls the FWER and offers greatly improved power compared with the non-hierarchical method; this improvement becomes especially noticeable as the number of experiments, $M$, increases. 
Figure~\ref{fig:HT_plot2} in Appendix~\ref{sec:addition} also indicates that our method continues to control the FWER with misspecified networks similarity and gives improved power; however, the improvement is less noticeable when the hierarchy is poorly constructed.

%%%%%%%%%%%%%%

%%%%%%%%%%%%%
\section{Application}\label{sec:data}
%%%%%%%%%%%%%

We consider the task of learning the functional connectivity network among a population of neurons, using the spike train data from \citet{Boldingeaat6904}.
In this experiment, spike times are recorded at 30 kHz on a region of the mice olfactory bulb (OB), while a laser pulse is applied directly on the OB cells of the subject mouse. The laser pulse is applied at increasing intensities at 8 levels from 0 to 50 ($mW/mm^2$). The laser pulse at each intensity level lasts 10 seconds and is repeated 10 times on the same set of neuron cells of the subject mouse. 

While a total of 80 laser stimuli were applied on neurons of multiple mice, for illustration purposes, we consider the spike train data collected at three stimuli at 0, 10 and 20 $mW/mm^2$ in a single mouse with the most neurons detected in OB ($p=25$ neurons). 
Since one laser pulse spans 10 seconds and the spike train data is recorded at 30 kHz, there are 300,000 time points per stimulus. We apply our joint estimation procedure using data under the three stimuli and evaluate the uncertainty of estimates using the hierarchical testing procedure. 

\begin{figure}
	\centering
	\includegraphics[width=1\linewidth]{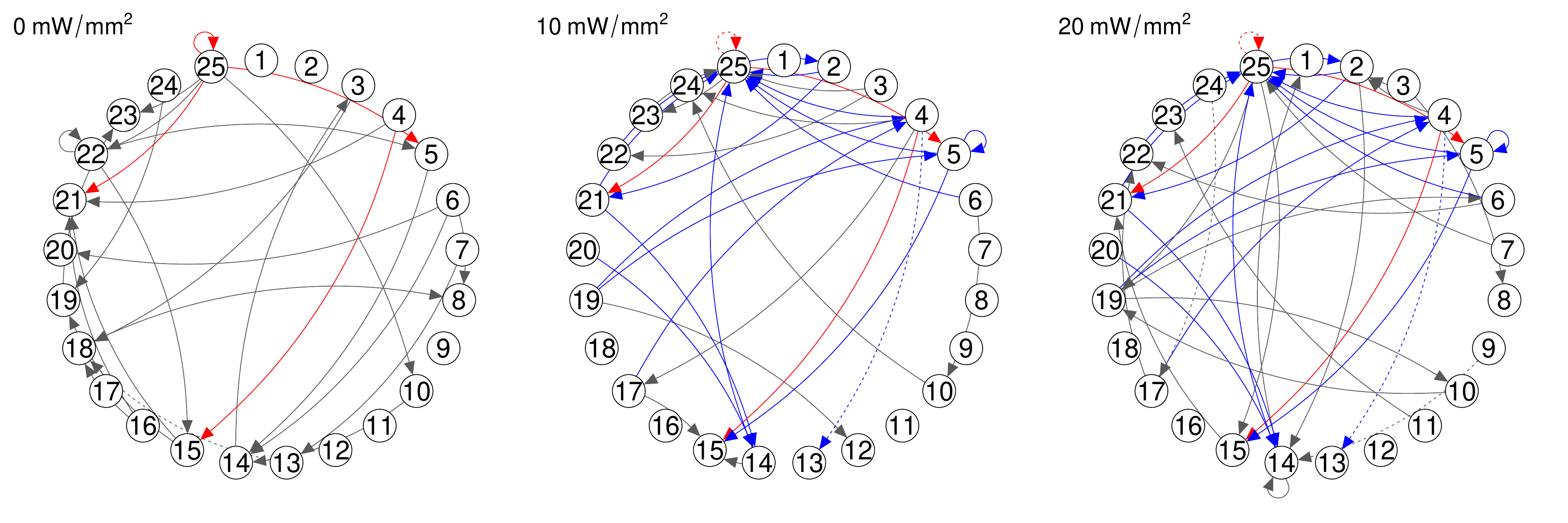}
	\caption{Estimated functional connectivities among neuronal populations using the spike train data from \citet{Boldingeaat6904}. Common edges across all experiments are in red. Edges shared only under laser conditions are in blue. Statistically significant edges, controlling FWER at $\alpha=0.05$ are shown in dashed lines. Edges that are unique to each condition are in gray. 
% 	\as{the figure is a bit too busy. pls remove gold edges and use two different line types to denote significant and non-significant edges; also why are there some dashed lines?} \sw{in the past we use gold edges for significant edges. To simplify the colors, I use dashed lines for the significant edges.}
	}
	\label{fig:dataexample}
\end{figure}

Figure~\ref{fig:dataexample} illustrates the estimated connectivity coefficients that are specific to each laser condition in a graph representation, where each node represents a neuron and a directed edge indicates a non-zero estimated connectivity coefficient.
More edges are observed when laser is applied (32 under 10 $mW/mm^2$ and 39 under 20 $mW/mm^2$ versus 27 under no laser). Both positive and negative edges are found in all conditions, corroborating the neuroscience hypothesis that both excitatory and inhibitory synapses facilitate maintaining stimulus specificity across odorant concentrations \citep{Boldingeaat6904}. Additionally, we find more common edges in the two laser conditions. Specifically, there are 17 edges (in blue) uniquely shared in the laser conditions compared to 4 edges (in red) shared in all conditions. To assess whether this difference is statistically significant, we generated randomly-connected networks with the same degree distributions at each of the three conditions and compared the observed difference to the distribution of the number of edges uniquely shared in the laser condition. This network permutation test indicates that the observed difference is unlikely under randomly generated networks ($p$-value $< 1e-5$).
This finding agrees with the observation by neuroscientists that the OB response is sensitive to the intensity level of the external stimuli \citep{Boldingeaat6904}. 

% \as{can we expand the discussion of findings based on \citet{Boldingeaat6904}? eg is there something more we can say about the patterns? We may also want to run a naive test comparing the number of common edges in the laser conditions with those common in all conditions -- for this we can assume independent draws of edges (so just generating networks with the same number of edges randomly). A slightly more sophisticated version would be to generate networks randomly with the same degree distribution (there are functions for this) and counting the number of common edges}
% \sw{
% i) will revisit \citet{Boldingeaat6904} and see any more scientific findings/interpretation corresponds to the patterns? 
% ii) permutation test, for each network, generate random random with the same number of edges and the degree need to be the same dist. as the observed one. Then, calcaulte the dffierence in num of edges common in laser and num of edges common in all, then decide the percentile. 
% Permutation test suggests a significant difference between the number of edges shared under the laser conditions and that shared over all conditions ($p$-value $< 1e-5$).
% }

%%

%%%%%%%%%%%%%
\section{Discussion}\label{sec:disc}
%%%%%%%%%%%%%
% summary
In this paper, we developed a joint estimation procedure for networks of high-dimensional Hawkes processes under multiple experiments. The optimization problem corresponding to our proposed estimation procedure is solved using a smoothing proximal gradient descent algorithm \citep{chenchen2008}. Although the algorithm works well for linear models, it empirically shows slow and unstable behavior with non-linear Hawkes models. Since non-linear link functions are often used when analyzing spike train data \citep{PANINSKI2007,Pillow2008}, developing more  computationally-efficient and stable algorithms for the non-linear models would be a potential direction of future research.  

% FDR control procedure 
Our proposed hierarchical testing procedure improves the testing power by taking advantage of the multi-experiment structure, while controlling the family-wise error rate. Given large-scale networks, a testing procedure that instead controls the FDR \citep[e.g.][]{BH1995} may offer additional power. 
\citet{Bogomolov2020} recently proposed a multiple testing adjustment procedure that controls the FDR by using the tree structure of the tests. While improving the power, the method requires a bottom-up $p$-value calculation, where the upper-level $p$-values needs to be a specific combination of those from the lower levels. Since all the hypotheses on the leaves need to be tested at beginning, such a procedure would be computationally intensive, particularly when the number of tests is large.
It is thus desirable to develop a procedure that allows $p$-values flexibly calculated on each node of the tree. For example, a procedure that allows a top-down %\as{I am not sure if this is the best word, do you mean top-down? or something else?} 
$p$-value calculation avoids intensive computation in calculating $p$-values over all the leaves, which becomes particularly important in sparse networks, when most leaf-hypotheses are null. Moreover, the existing literature \citep[e.g.][]{Li2019, Bogomolov2020} that control FDR for structured tests often require the structure to be correctly identified. Given that such structural information is not always available or may not be accurately estimated, developing FDR controlling procedures that are robust to the structure misspecification would be another direction of future research.

% draft: discussion on limitation of weights based on shared correlation edges 
Theorem~\ref{fwer_general} shows that, for large and sparse networks, our proposed hierarchical testing procedure is robust to potential misspecification of the hierarchical structure defined based on the proposed similarity in \eqref{def:cv_dist}. Nonetheless, consistent estimation of similarities between networks may still be of interest. In particular, such an estimate could, for instance, facilitate the development of hierarchical FDR controlling procedures discussed above. 
A key requirement for developing consistent estimates of similarities between connectivity networks based on cross-covariances is to develop a measures of similarity that is \textit{order-preserving}; that is, the order of the similarity based on cross-covariances is the same as that given by the oracle similarity based on the (unknown) connectivity network. 
% However, our proposed network similarity in \eqref{def:cv_dist} is only heuristic. There is no guarantee that this similarity is \textit{order-preserving}; that is, the order of the similarity based on the number of shared non-zero cross-covariances is the same as that given by the oracle similarity based on the true edges. 
One such measure of similarity can be defined based on the connected components of the networks. A \textit{connected component} is a set of nodes that are connected by paths in an undirected graph. In the setting of our problem, the  edges of this undirected graph are given by
$ \mathcal{E}^{u} = \left \{ (i,j):  | \beta_{ij} |  \ne 0 , 1\le i, j\le p \right \} $. 
%The connect component essentially defines a clustering of the network nodes. 
Let $\{ \mathcal{C}^{(m)}_l \}_{l=1}^{L^{(m)}}$ and $\{ \mathcal{C}^{(m')}_l \}_{l=1}^{L^{(m')}}$ denote the  connected components of two $p$-variate networks in conditions $m$ and $m'$. The \textit{connected-component similarity} can be defined as 
\begin{align}
    d^{cc}\left ( \{ \mathcal{C}^{(m)}_l \}_{l=1}^{L^{(m)}}, \{ \mathcal{C}^{(m')}_l \}_{l=1}^{L^{(m')}} \right) 
    = \sum_{l,l'} r_{l,l'} \left \{ \log ( r_{l,l'}/p_{l} ) +    \log ( r_{l,l'}/q_{l'} ) \right \} , 
\end{align}
where $ p_l   = \big\lVert \mathcal{C}^{(m)}_l  \big\rVert  / p$,  
    $q_{l'}   = \big\lVert \mathcal{C}^{(m')}_l  \big\rVert  / p $ 
    and 
    $ r_{l,l'} = \big\lVert \mathcal{C}^{(m)}_l  \cap \mathcal{C}^{(m')}_{l'}  \big\rVert /p $.
This measure, which is also known as \textit{variation of information}, is often used to compare the similarity between two clusterings \citep{marina2003}. 

While the true connected components are unknown in practice, they can be consistently estimated. In particular, we can obtain estimates $\{ \widehat{\mathcal{C}}^{(m)}_l  \}_{l=1}^{\hat{L}^{(m)}}$ from undirected graphs corresponding to the nonzero values of thresholded empirical cross-covariances (with thresholding at $\kappa$) as 
$ \widehat{\mathcal{E}}^{u}(\kappa) = \left \{ (i,j):  \big| \widehat{V}_{ij} \big| > \kappa, 1\le i, j\le p  \right \} $. 
\citet{shizhe_thesis_2016} has shown that 
%\as{state the exact result} 
the connected components of the true network can be consistently identified using the empirical cross-covariances---i.e., 
$\mathbb{P}\left ( \{ \mathcal{C}^{(m)}_l \}_{l=1}^L  =\{ \widehat{\mathcal{C}}^{(m)}_l  \}_{l=1}^{\hat{L}^{(m)}} \right) \rightarrow  1$ as $T^{(m)} \rightarrow \infty$ with $\kappa= o\left ( \left( T^{(m)}\right)^{-1/5}\right)$. 
Thus, as a natural estimator of $d^{cc}\left ( \{ \mathcal{C}^{(m)}_l \}_{l=1}^{L^{(m)}}, \{ \mathcal{C}^{(m')}_l \}_{l=1}^{L^{(m')}} \right)$, 
$d^{cc}\left ( \{ \widehat{\mathcal{C}}^{(m)}_l \}_{l=1}^{\widehat{L}^{(m)}}, \{ \widehat{\mathcal{C}}^{(m')}_l \}_{l=1}^{\widehat{L}^{(m')}} \right)$, based on the empirical cross-covariances, consistently represent the similarity in the connected components of the true networks, thus it is order-preserving.
% by continuous mapping theorem where $d^{cc}$ is a continuous function of $p, q, r$. Thus, $\hat{d}^{cc} \rightarrow d^{cc}$, which means asymptotically order-preserving.
%
% It can be shown that if $\widehat{d}^{cc}$ is a consistent estimate of ${d}^{cc}$ the empirical connected-component similarity, $\widehat{d}^{cc}$ , is order-preserving. \as{perhaps clarify why} 
% However, this consistency follows from a result in \citet{shizhe_thesis_2016} \as{state the exact result} \sw{it seems it is stated as follows?}, which shows that, with a proper threshold, the connected components of the true network can be consistently identified using the empirical cross-covariances; formally, 
% $\mathbb{P}\left ( \{ \mathcal{C}^{(m)}_l \}_{l=1}^L  =\{ \widehat{\mathcal{C}}^{(m)}_l(\eta) \}_{l=1}^{\hat{L}^{(m)}} \right) \rightarrow  1$ as $T^{(m)} \rightarrow \infty$ with $\kappa= o\left ( \left( T^{(m)}\right)^{-1/5}\right)$. 
%Therefore, similarities among connected components can be used as a consistent measure of similarity among networks. 
However, the effectiveness of a similarity measure based on connected-components depends on the structure of the underlying networks. For instance, while the networks of circles and stars in Figure~\ref{fig:network_star_circle} are quite different, their connected-component structures are identical. As a result, similarity weights and dendrograms defined based on connected component may not be informative in this case. 
% (see Figure~\ref{fig:network_SE_CC} in Appendix~\ref{sec:addition}). 
Developing more effective  order-preserving network similarity measures is thus an important area of future research.

\bibliography{ref}
\clearpage 
\appendix
%%%%%%%%%%%%%
\section{The Smoothing Gradient Descent Algorithm}\label{sec:algo}
%%%%%%%%%%%%% 

%The fusion penalty makes \eqref{eq:optimization} challenging to solve \citep{friedman2007}. 
Our estimator in \eqref{eq:optimization} is solved using the smoothing proximal gradient descent algorithm \citep{ChenSPGD2012}. The algorithm replaces the non-smooth fusion penalty by a smoothing approximation thus makes the original problem easier to solve using the fast iterative shrinkage thresholding algorithm \citep{Beck2009AFI}. In the follows, we specify the algorithm to the cases of linear and non-linear Hawkes processes.

Let $I_0 = \mat{0 & 0 \\
	0 & I_p} \in \mathbb{R}^{(p+1)\times (p+1)}$ 
and $I_p$ is the identity matrix. 
%In addition, let $	\bm{\theta}_i= \left(  \left( \bm{\theta}_i^{(1)} \right)^\top, \dots, \left( \bm{\theta}_i^{(M)} \right)^\top \right)^\top $. 
Then the sparse penalty is written as
\[
\lambda_1 \sum_{m=1}^M  \lVert \bm{\beta}^{(m)}_i \rVert_1
= \lVert \Lambda \bm{\theta}_i \rVert_1 ,
\]
where $\Lambda = 
\lambda_1 I_M \otimes I_0$ and $I_M \in \mathbb{R}^{M \times M}$ is an identity matrix. 

%For $1\le k < l\le p$, we write 
%\begin{align}
%\lambda_2 w_{k,l}\sum_{1\le i \le p} \left \lVert \bm{\beta}^{(k)}_i - \bm{\beta}^{(l)}_i \right \rVert_1
%=  \left \lVert \bm{c}^\top_{k,l} \bm{\theta}_i \right \rVert_1, 
%\end{align}
%where $\bm{c}_{k,l} = \lambda_2 w_{k,l} (\bm{e}_k - \bm{e}_l) \otimes (0,\mathbf{1}_{p}^\top) \in \mathbb{R}^{(p+1)M}$ and $\bm{e}_k$ is canonical basis in $\mathbb{R}^p$.
%Define $\bm{d}_k = (\bm{d}_{k,k+1},\dots,\bm{d}_{k,K})^\top$, where $\bm{d}_{k,l} = w_{k,l} (\bm{e}_k - \bm{e}_l) \otimes (0,\mathbf{1}_{p}^\top) \in \mathbb{R}^{(p+1)K}$. Let $D = (\bm{d}_1,\dots, \bm{d}_{K-1})^\top \in \mathbb{R}^{ {K \choose 2}\times (p+1)K }$, and $C = \lambda_2 D$. 
%
Let $d_{m,m'} = w_{m,m'} (\bm{e}^\top_{m} - \bm{e}^\top_{m'}) \otimes I_0 \in \mathbb{R}^{(p+1)\times (p+1)M}$, where $\bm{e}_m$ is canonical basis in $\mathbb{R}^M$.
Let $D = \mat{ d_{1,2} \\
                 \dots \\
                 d_{m,m'} \\
                \dots \\ d_{M-1, M}  } \in \mathbb{R}^{ {M \choose 2}(p+1) \times M (p+1)}$. 
Define $C = \lambda_2 D$. Then the fusion penalty becomes
\begin{align}
\lambda_2   \sum_{1\le m < m' \le M} w_{m,m'}  \left \lVert \bm{\beta}^{(m)}_i - \bm{\beta}^{(m')}_i \right \rVert_1
= \left \lVert C \bm{\theta} \right \rVert_1.
\end{align} 

Thus, the penalty in \eqref{eq:penalty} can be written as  
\begin{align}
\mathcal{P}( \bm{\theta}_i  ) 
&= \lVert \Lambda \bm{\theta}_i \rVert_1 
+   \lVert C \bm{\theta}_i \rVert_1, 
\end{align}
where the fusion penalty can be written using its dual norm as
\begin{align}
\left \lVert C \bm{\theta}_i  \right \rVert_1 
 =  \max_{  \lVert \bm{\alpha}_i \rVert_{\infty} \le 1 } \bm{\alpha}_i^T C \bm{\theta}_i .
\end{align}
Next, let $f_u(\bm{\theta}_i) $ be a smoothing approximation function such that   
\begin{align}
f_u(\bm{\theta}_i) 
=  \left \lVert C \bm{\theta}_i  \right \rVert_1
- u \frac{1}{2} \left  \lVert \bm{\alpha}_i  \right \rVert_2^2 
= \max_{\lVert \bm{\alpha}_i \rVert_{\infty} \le 1 }  
\bm{\alpha}_i^T C  \bm{\theta}_i   - u \frac{1}{2}  \lVert \bm{\alpha}_i \rVert_2^2
 \label{smooth_function} ,
\end{align}
where $u > 0$ is a smoothness parameter that controls the level of approximation to the fusion penalty. 
For example, if we take $u = \frac{4 \epsilon}{ M(M-1)}$, then the approximation error is up to $\epsilon$. 
\citet{ChenSPGD2012} shows that 
$f_u (\bm{\theta}_i)$ is convex and continuous differentiable where  
\begin{align}
\label{eq:df_u}
\nabla f_u (\bm{\theta}_i) = C^T \bm{\alpha}_i^* .
\end{align}
Here $\bm{\alpha}_i^* = S( \frac{C\bm{\theta}_i }{u})$ and $S(\bm{z})$ is a coordinate-wise projection operator that projects each entry of $\bm{z}$ to the $\ell_\infty$-ball --- i.e.,
\begin{align*}
S(z) = 
\begin{cases}
-1 & z < 1 \\  
z &  z\in [-1,1] \\
1 & z > 1  \\ 
\end{cases} .
\end{align*}

Substituting the fusion penalty with \eqref{smooth_function},  solving \eqref{eq:optimization} becomes to solve
\begin{align}
\widehat{\bm{\theta}}_i 
& = \arg \min_{
	\bm{\theta}_i \in \mathbb{R}^{ (p+1)M} }
\left \{  
h(\bm{\theta}_i) +  \lVert \Lambda_i \bm{\theta}_i \rVert_1 
\right \}  \quad i=1,\dots, p ,
\label{eq:optimization_spgd}
\end{align}
where 
\[
h(\bm{\theta}_i) =
\frac{1}{T }
\sum_{m=1}^M
\int_{0}^{T_m}
\ell \left (   dN^{(m)}_i(t), f_{\bm{\theta}^{(m)}_i }(\bm{x}^{(m)}(t)) \right )  
+
f_u(\bm{\theta}_i)
,
\]
and $T = \sum_{m=1}^M T_m$. 
Problem~\eqref{eq:optimization_spgd} can be solved using the fast iterative shrinkage thresholding (FISTA) algorithm \citep{Beck2009AFI}. The FISTA algorithm is a first-order optimization method \citep{Beck2017} that requires evaluating the first derivative of $h(\bm{\theta}_i)$, and a Lipschitz constant $L$ calculated as a upper bound on the spectral radius of the second derivative of $h(\bm{\theta}_i)$. 

Next, we specify the first derivative of $h(\bm{\theta}_i)$ and
the Lipschitz constant $L$ for the linear Hawkes model with least square loss --- i.e., $\ell(a,b) = (a-b)^2 $. 

Let
$\gamma^{(m)}_{ij} =  \int_0^{T_m}   x^{(m)}_j(t) dN^{(m)}_i(t) $,
$\gamma^{(m)}_{i0} =   \int_0^{T_m}    dN^{(m)}_i(t) $, and $\bm{\gamma}^{(m)}_i = \left (\gamma^{(m)}_{i0},\gamma^{(m)}_{i1},\dots,\gamma^{(m)}_{ip} \right )^\top  \in \mathbb{R}^{p+1}$. 
Denote $\textrm{Q}^{(m)}=  \int_0^{T_m}   \mat{1 \\ \bm{x}^{(m)}(t) } 
\mat{1 &  \left( \bm{x}^{(m)}(t) \right)^\top } dt$. Let 
$
\textrm{Q} = \mat{ \textrm{Q}^{(1)} &         &          & \\ 
	& \textrm{Q}^{(2)} &          & \\  
	&         &  \dots   & \\  
	&         &          & \textrm{Q}^{(M)} 
} ,
$
and $\bm{\gamma} =\left (  \left( \bm{\gamma}^{(1)} \right)^\top , \dots ,  \left( \bm{\gamma}^{(m)} \right)^\top  \right )^\top  $, where $\textrm{Q}^{(m)}$ and $\bm{\gamma}^{(m)}_i$ can be pre-calculated given data and the pre-specified transfer kernel function. With these notations, 
\begin{align}
\ \sum_{m=1}^M
\int_{0}^{T_m}
\ell \left (   dN^{(m)}_i(t), f_{\bm{\theta}^{(m)}_i }(\bm{x}^{(m)}(t)) \right ) 
= 
  \bm{\theta}_i^\top  \textrm{Q} \bm{\theta}_i
- 2  \bm{\theta}_i^\top  \bm{\gamma}_i  + \sum_{m=1}^M \int_0^{T_m}  \left( dN^{(m)}_i(t) \right)^2,
\label{eq:ls_loss}
\end{align}
which leads to
\begin{align} 
\nabla h(\bm{\theta}_i) &=  
\frac{1}{T}\textrm{Q}\bm{\theta}_i - \frac{1}{T}\bm{\gamma}  + C^T \bm{\alpha}^*   .
\label{eq:nablah_ls}
\end{align}
In addition, $\nabla^2 h(\bm{\theta}_i)  \preccurlyeq \frac{1}{T} \textrm{Q}  +  \frac{C^TC}{u}  $, which leads us to choose $L = \Lambda_{\max}\{ \frac{1}{T}\textrm{Q} + \frac{C^TC}{u} \}$. Notice that both $\nabla h(\bm{\theta}_i)$ and $L$ can be pre-calculated and stored in memory when implementing the FISTA algorithm. This feature makes the computation scalable to large size data collected over a long time range. The edge selection performance using this algorithm for the linear Hawkes process is illustrated in Figure~\ref{fig:lstpfp2} in Section~\ref{sec:sims}. 

Next, we specific $\nabla h(\bm{\theta}_i)$ and $L$ for non-linear Hawkes model with the exponential-link function, $g_i(\cdot) = \exp(\cdot)$ and the negative log likelihood loss --- i.e., 
$\ell \left ( a,  b \right )  = - a \log (b)  + b $.

With some algebra, 
\begin{align}\label{eq:nablah_glm}
\nabla h(\bm{\theta}_i) &= 
\frac{1}{T} \widetilde{Q}_i   -  \frac{1}{T} \bm{\gamma}   +  C^T \bm{\alpha}^* , 
\end{align}
where 
\begin{align*}
\widetilde{Q}_i
= 
\mat{ \widetilde{Q}^{(1)}_i  &         &          & \\ 
	& \widetilde{Q}^{(2)}_i    &          & \\  
	&         &  \dots   & \\  
	&         &          & \widetilde{Q}^{(M)}_i  
}  &, \\
\widetilde{Q}^{(m)}_i
= \int_0^{T_m} \bm{x}^{(m)}(t) 
\exp \left(   
\mat{ 1 & \left(\bm{x}^{(m)}(t)\right)^\top}  \bm{\theta}^{(m)}_i 
\right)  dt &, \quad m \in \{1,\dots, M \}  .
\end{align*}
Unlike the linear model, $\nabla h(\bm{\theta}_i)$ depends on the unknown value of the parameter. Therefore, we need to evaluate $\nabla h(\bm{\theta})$ at each step in the FISTA algorithm, which slows down the algorithm given observations over long time periods.

\begin{figure} 
	\centering
	\includegraphics[width=0.6\linewidth]{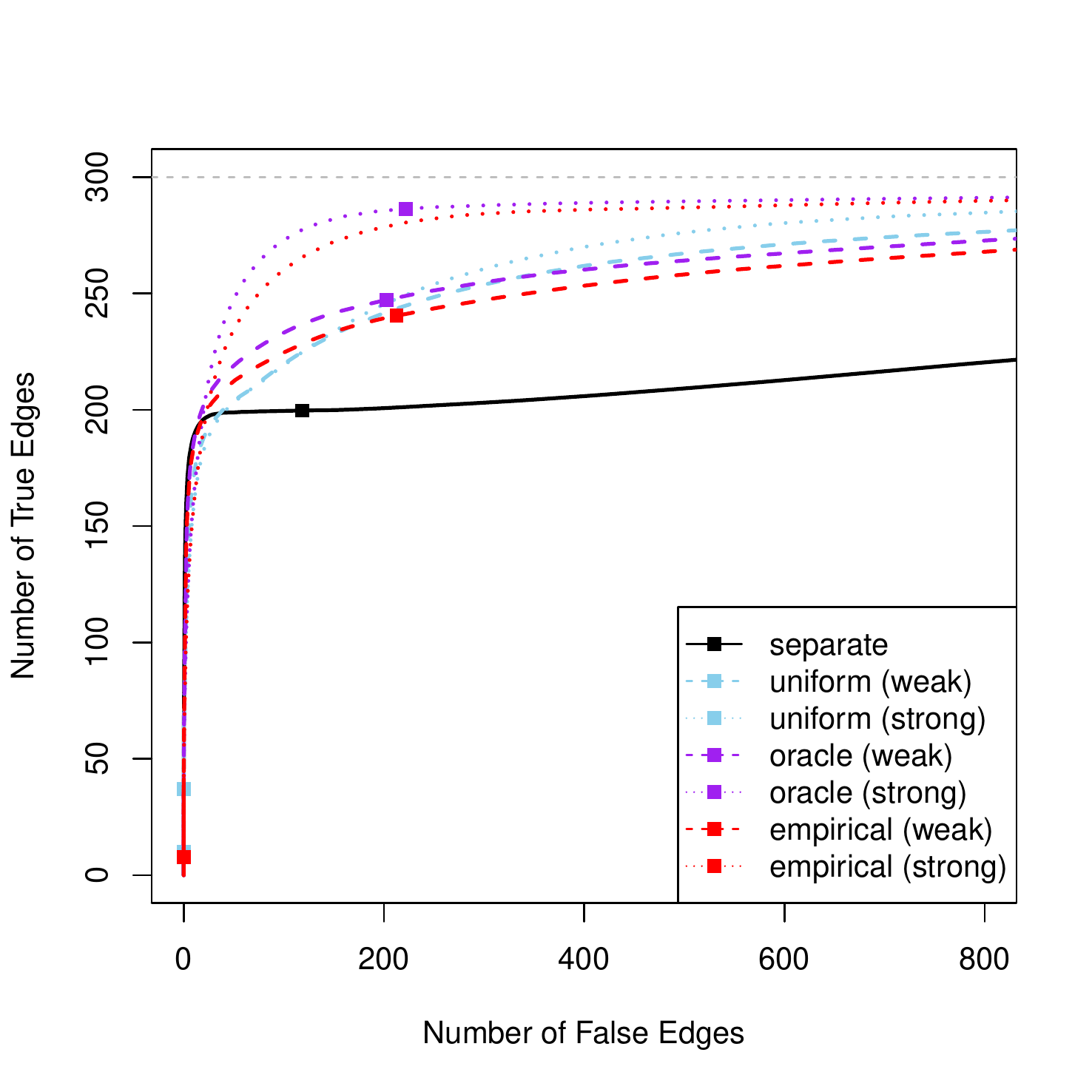}
	\caption{
	Edge selection performance of the proposed joint estimation method in a simulation study focused on inferring edges in 3 networks of generalized Hawkes processes with exponential link function. The plots show average number of true positive and false positive edges, over 100 simulation runs, for the joint estimation method with different choices of weights, compared to separate estimation of each network. Weight strategies include oracle, empirical and uniform weights. Solid squares ($\blacksquare$) correspond the choice of tuning parameter using eBIC. 
	%(a): Network~2 shares 90\% edges with Network 1 and 10\% with Network 3 as in Figure~\ref{fig:network_star_circle}. (b): Network 2 is the same as Network~1. 
	%\as{update caption based on the edits in the main paper} 
	%\sw{updated} 
	}
	\label{fig:glmtpfp}
\end{figure}

Notice that 
$
\nabla^2 h(\bm{\theta}_i) \preccurlyeq 
\max_{ 1\le m \le M}\left \{ 
\exp
\left ( 
\mat{ 1 & \left(\bm{x}^{(m)}(t)\right)^\top}  \bm{\theta}^{(m)}_i 
\right  )
\right \} \frac{1}{T}\textrm{Q} +\frac{C^TC}{u}
$ ,
which leads to a choice of $L$. However, we find that this choice of $L$ leads to slow convergence or even divergence when $\frac{C^TC}{u}$ is large --- e.g., with large $\lambda_2$ and small $u$. 
To mitigate this issue, we use a general convex programming solver as an alternative --- e.g., \texttt{CVXR} in \texttt{R} \citep{cvxr2020} ---when the algorithm meets convergence problem.
%\as{what does this mean?} \sw{when the SPGD algorithm does not converge, I use CVXR to solve the optimization problems}. 
The edge selection performance using the algorithm for the non-linear Hawkes process is illustrated in Figure~\ref{fig:glmtpfp}. 

%%%%

We summarize the computational steps described above in Algorithm~\ref{alg:spgd}. We note that when the number of experiments, $M$, is large, $u$ becomes very small (proportional to $O(1/M^2)$), which may lead to very large $L$. In that case, the algorithm may converge slowly, because the step-size, $1/L$, becomes very small.  

%Then, we repeat the algorithm separately for all the $p$ component processes to obtain the estimates on $\{\bm{\theta}_i \}_{i=1}^p$. 

\begin{algorithm}[h] 
    \caption{Smoothing Proximal Gradient Descent for Generalized Hawkes Process}
	\begin{algorithmic} \label{alg:spgd}
	\FOR {$i=1,\dots, p$}
		\STATE {Input: $\{N^{(m)}_i\}_{m=1}^M$, $C$, 			           
			           $\Lambda$,
			           $\bm{\theta}_i^0$, 
			           $L$, desired accuracy $\epsilon$  }
		\STATE{Initialization: set 
			$u = \frac{4\epsilon}{(M-1)M}$, 
			$\delta^0= 1$, $\bm{w}^0 = \bm{\theta}_i^0$ }		
		\REPEAT 
		\STATE {1: Compute $\nabla h(\bm{w}^t)$ ; }
		\STATE {2: Solve the proximal operator associated with $\ell_1$-norm penalty:
			\begin{align*}
			\bm{\theta}_i^{t+1} &= 
			\arg \min_{\bm{\theta}_i } 
			\left (  
			h(\bm{w}^t) + (\bm{\theta}_i - \bm{w}^t)^\top \nabla h(\bm{w}^t) 
			+ \frac{L}{2} \lVert \bm{\theta}_i - \bm{w}^t \rVert_2^2
			+ \lambda_1 \lVert \bm{\theta}_i \rVert_1
			\right)	\\
			&= \arg\min_{ \bm{\theta}_i } 
			\frac{1}{2} \lVert \bm{\theta}_i - \bm{v} \rVert_2^2  + 
			\frac{\lambda_1}{2} \lVert \bm{\theta}_i  \rVert_1 \\  
			&= sign(\bm{v}) \max\left(  0, |\bm{v}| - 
			\frac{ diag(\Lambda) }{L} \right),
			\end{align*}
			where $\bm{v}  = \bm{w}^t - \frac{1}{L} \nabla h(\bm{w}^t)$.				
		}
		\STATE{3: $\delta^{t+1} = \frac{2}{t +3 }$ ; } 
		\STATE {4: $\bm{w}^{t+1} = \bm{\theta}_i^{t+1} + \frac{1-\delta^t}{\delta^t} \delta^{t+1} (\bm{\theta}_i^{t+1}- \bm{\theta}_i^t)$ }	
		\UNTIL {convergence of $\bm{\theta}_i^t$ ;}
	\ENDFOR
	\end{algorithmic}
\end{algorithm}

% another setting when Network 2 is not exact the same as Network 1 
%We run additional simulation where the networks of the first two experiments are similar but not exactly the same (Figure~\ref{fig:network}). While the joint estimation procedure gives favorable edge selection performance, its effectiveness is less pronounced because the first two networks are set to be less similar than before (Figure~\ref{fig:ls_tpfp}).

\clearpage
%%%%%%%%%%%%%%%%%%%%%%%%%%%%%%%%%%%%%%%%%%%%%%%%%%%%%
\section{Proofs}\label{sec:proofs}
% define notations 
For ease of presentation, we stack the data from $M$ experiments and re-label the time index over $[0, T]$ where $T= \sum_{m=1}^M T_m$. 
In particular, we denote $N_i(t)=N_i^{(m)}(t-\sum_{l=0}^{m-1} T_l )$ if $t\in  \big(\sum_{l=0}^{m-1} T_l, \sum_{l=0}^m T_l  \big ]$ for $1\le m \le M$, where $T_0 =0$. Then,
denote
$$\bm{x}(t) = 
\mat{ \mathbf{1}(T_0 < t \le T_1) \\  
	\bm{x}^{(1)}(t)  \mathbf{1}( T_0< t \le T_1) \\ \dots \\ 
	\mathbf{1}(\sum_{m=0}^{M-1} T_m< t \le T) \\
	  \bm{x}^{(M)}(t) \mathbf{1}(\sum_{m=0}^{M-1} T_m< t \le T)
      },
  \quad  
  \bm{\theta}_i = \mat{ \bm{\theta}_i^{(1)} \\
  	 \dots \\
  	   \bm{\theta}_i^{(M)} } , 
     \quad 
     \lambda_i(t) = g_i( \bm{x}^\top(t) \bm{\theta}_i ) .
 $$
With these notations, 
\begin{align*}
\sum_{m=1}^M
\int_{0}^{T_m}
\ell \left (   dN^{(m)}_i(t), f_{\bm{\theta}^{(m)}_i }(\bm{x}^{(m)}(t)) \right )  
= \int_{0}^{T}
\ell \left (   dN_i(t), f_{\bm{\theta}_i }(\bm{x}(t)) \right ) .
\end{align*}
In addition, denote $\ell( t;\bm{\theta}_i ) = \ell \left (   dN_i(t), f_{\bm{\theta}_i }(\bm{x}(t)) \right ) $. 
%
%Recall the following definitions introduced in Section~\ref{sec:theory}.
%Define the set of active indices, $S^{(m)}_i= \{j:  \beta^{(m)}_{ij}\ne 0 , 1\le j \le p \}$, and $d^{(m)}_i = |S^{(m)}_i|$ and $d^* \equiv \max_{ 1\le m\le M, 1\le i \le p} d_i^{(m)} $, and $S_i = \bigcup_{m=1}^M S_i^{(m)}$. 
%Let the set of dissimilar experiment indices, $\widetilde{S}_i= \{ (j,m): \beta_{ij}^{(m)}\ne \beta_{ij}^{(m')},  \exists  m'\ne m \in \{1 ,\dots, M\}, \text{ for } 1\le j \le p \}$. 
%It is easy to see that $|S_i|\le Md^*$ and $ r^* \equiv \max_{1\le i \le p} |\widetilde{S}_i|  \le Md^*$. In addition, let $\Delta_i^{(m)} = \widehat{\bm{\theta} }_i^{(m)}  - \bm{\theta}_i^{(m)}$, and $\Delta_i = \left ( \left( \Delta_i^{(1)}  \right)^\top, \dots, \left( \Delta_i^{(M)} \right)^\top \right )^\top \in R^{(p+1)M} $. With these notations, $\Delta_{S}$ and $\Delta_{\widetilde{S} }$ are vectors that collect the estimation error for all non-zero $\beta_{ij}^{(m)}$, and all $\beta_{ij}^{(m)}$ that does not have the same value across experiments, respectively. 

Because optimization problem \eqref{eq:optimization} can be solved separately for each component process, in the following we illustrate the estimation consistency using the estimator \eqref{eq:optimization} for one component process. Moreover, for ease of notation, we drop the subscript $i$; that is, we use $\bm{x}(t)$ for $\bm{x}_i(t)$, $\bm{\theta}$ for $\bm{\theta}_i$, $dN(t)$ for $dN_i(t)$, $g(\cdot)$ for $g_i(\cdot)$ and $\lambda(t)$ for $\lambda_i(t)$, $S$ for $S_i$, $\widetilde{S}$ for $\widetilde{S}_i$ and $A$ for $A_i$. 

%\as{I am restructuring the appendix a bit, bringing the supplementary lemmas before the proofs and not separating proofs of theorems 1 and 2 from 3 and 4.}

Next, we state two lemmas needed for the proofs of Theorems~\ref{theorem1} and \ref{theorem2}. 

\begin{lemma}[\citet{vandegeer1995}]
	\label{lemma_vandergeer1995}
	Suppose there exists $\lambda_{\max} $ such that $\lambda(t) \le \lambda_{\max}$ where  $\lambda(t)$ is the intensity function of Hawkes process defined in 
	\eqref{eq:intensity_prob}. Let $H(t)$ be a bounded function that is $\mathcal{H}_t$-predictable.
	Then, for any $\epsilon > 0$,
	\begin{align*}
	\frac{1}{T} \int_0^T H(t) \bigg \{ \lambda(t) dt - dN(t) \bigg \}
	\le 4 \bigg \{ \frac{\lambda_{\max} }{2T }   \int_0^T H^2(t) dt \bigg \}  ^{1/2} \epsilon^{1/2},
	\end{align*}
	with probability at least $1 - C\exp(-\epsilon T)$, for some constant $C$.
\end{lemma}

\begin{lemma}[\citet{wang2020statistical}]
\label{lemma_min_eigen}
Suppose the Hawkes process defined in \eqref{eq:hawkes_para_transfer}
satisfies Assumptions~\ref{assumption1}--~\ref{assumption4}. 
Let $\textrm{Q} =  \frac{1}{T}\int_0^{T}   \mat{1 \\ \bm{x}(t) } 
\mat{1 &  \bm{x}^\top(t)} dt$, where $\bm{x}(t)$ is defined in \eqref{eq:design_column_xt}. Then, there exists $\gamma >0$ such that
\begin{align*}
\Lambda_{\min}\left( \textrm{Q}  \right) \ge \gamma > 0,
\end{align*}
with probability at least $1-c_1 p^2 T \exp(-c_2 T^{1/5})$, where constants
$c_1, c_2$ depending on the model parameters and the transition kernel.
\end{lemma}

\hfill %\break

%%%%%%%%%%%%%%%%%%%%%%%%%%%%%%%%%
% proof of theorem 1
%%%%%%%%%%%%%%%%%%%%%%%%%%%%%%%%%%
\noindent
\textbf{Proof of Theorem~\ref{theorem1}}:

Let $\Delta  = \widehat{\bm{\theta}} - \bm{\theta}$.
We linearize $\ell( t;\bm{\theta} )$  w.r.t. $\bm{\theta}$ using Taylor expansion: 
\begin{align}
\ell( t;\widehat{\bm{\theta}} ) - \ell(t;\bm{\theta})
&= \left( \nabla\ell(t;\bm{\theta}) \right)^\top \Delta   +  \frac{1}{2} \Delta^\top \nabla^2\ell(t;\bm{\theta})   \Delta   + o( \lVert \Delta \rVert^2_2). \label{eq:talyor_expan}
\end{align}

Let
$R(\bm{\theta}) = \left \lVert \Lambda \bm{\theta} \right \rVert_1  + \left \lVert  C \bm{\theta} \right \rVert_1 $, where $\Lambda$ and $C$ are defined in Appendix~\ref{sec:algo}. Taking $ \widehat{\bm{\theta} }$ given by  \eqref{eq:optimization}, 
\begin{align}
\label{eq:ineq_from_optim}
\frac{1}{T}\int_0^{T} \left\{ \ell( t;\widehat{\bm{\theta}} ) 
- \ell(t;\bm{\theta}) \right \}  = 
\frac{1}{T}\int_0^{T} \ell( t;\widehat{\bm{\theta}} ) 
-\frac{1}{T}\int_0^{T}\ell(t;\bm{\theta}) 
\le  R(\bm{\theta} ) - R(  \widehat{\bm{\theta}}  ). 
\end{align}
Taking \eqref{eq:talyor_expan} in \eqref{eq:ineq_from_optim}, 
%\as{I am not quite following this part} \sw{hopefully this makes better}
\begin{align} \label{eq:basic_ineq}
0\le \Delta^\top \left( \frac{1}{2T} \int_0^{T}  \nabla^2\ell(t;\bm{\theta}) \right) \Delta  
\le -\frac{1}{T} \int_0^{T}  \left( \nabla\ell(t;\bm{\theta}) \right)^\top \Delta   + R(\bm{\theta} ) - R(  \widehat{\bm{\theta}} ).
\end{align}
% restriction on $\Delta$
%Before we proceed, notice that there is a constraint on $\Delta$ such that
%$\Delta \in \mathcal{C}= \{ \Delta \mid  4 \lVert \Delta_S \rVert_1  \ge 
%\lVert \Delta_{S^c} \rVert_1  \} $ because
%\begin{align*}
%0 &\le  -\frac{1}{T} \sum_{m=1}^M \nabla\ell(t;\bm{\theta}) \Delta   + R(\bm{\theta}^{(m)} ) - R(  \widehat{\bm{\theta}}^{(m)} ) \\
%& \le
%- \nabla\ell(t;\bm{\theta}) \Delta 
%+ \rho_1 \lVert \bm{\theta} \rVert_1
%- \rho_1 \lVert \widehat{\bm{\theta} } \rVert_1
%+ \rho_2 \lVert  C\bm{\theta} \rVert_1
%- \rho_2 \lVert  C\widehat{\bm{\theta} } \rVert_1 \\
%&\le 
%\frac{1}{2} \rho_1 \lVert \Delta_S \rVert_1  
%+ \frac{1}{2} \rho_1 \lVert \Delta_{S^c} \rVert_1 
%+ \rho_1 \lVert \Delta_{S} \rVert_1
%- \rho_1 \lVert \Delta_{S^c} \rVert_1
%+ \frac{1}{2}\rho_1  \lVert \Delta_{S} \rVert_1 \\
%& \le   2 \rho_1  \lVert \Delta_{S} \rVert_1  
%- \frac{1}{2} \rho_1 \lVert \Delta_{S^c} \rVert_1 .
%\end{align*}

Let  $A = S \cap \widetilde{S}^c$. Recalling the definition of $S$ and $\widetilde{S}$ in Section~\ref{sec:theory}, $A$ is the set of indices associated with the coefficients that are nonzero (i.e., in $S$) and have the same values under all conditions (i.e., in $  \widetilde{S}^c$). 
Then,
\begin{align*}
 R\left(\bm{\theta} \right) - R\left(  \widehat{\bm{\theta}}  \right) 
 &= 
 \rho_1 \lVert \bm{\theta} \rVert_1 
 + \rho_2 \lVert D \bm{\theta} \rVert_1 
 - 
 \rho_1 \lVert \widehat{\bm{\theta} } \rVert_1 
 -\rho_2 \lVert D \widehat{\bm{\theta} } \rVert_1 \\
 &= 
 \rho_1 \lVert \bm{\theta}_A \rVert_1 
 +\rho_1 \lVert \bm{\theta}_{S\cap \widetilde{S}} \rVert_1 
 - 
 \rho_1 \lVert \widehat{\bm{\theta}}_A  \rVert_1 
 -\rho_1 \lVert \widehat{\bm{\theta}}_{S\cap \widetilde{S} } \rVert_1 
 -\rho_1 \lVert \widehat{\bm{\theta}}_{S^c}  \rVert_1 
 \\
 &\quad + \rho_2 \lVert D \bm{\theta} \rVert_1 
 - \rho_2 \lVert D \widehat{\bm{\theta} } \rVert_1 \\
 & \le 
 \rho_1 \lVert  \Delta_A \rVert_1   + 
 \rho_1 \lVert  \Delta_{ S \cap \widetilde{S}} \rVert_1    
 -\rho_1 \lVert  \Delta_{S^c} \rVert_1  
 + \rho_2 \left  \lVert D_{., \widetilde{S} } \Delta_{  \widetilde{S} } \right \rVert_1 
 - \rho_2 \left  \lVert D_{.,\widetilde{S}^c} \Delta_{ \widetilde{S}^c} \right \rVert_1 \\
 &=
 \rho_1 \lVert  \Delta_S \rVert_1 
 -\rho_1 \lVert  \Delta_{S^c} \rVert_1  
 + 
 \rho_2 \left  \lVert D_{., \widetilde{S} } \Delta_{  \widetilde{S} } \right \rVert_1 
 - \rho_2 \left  \lVert D_{.,\widetilde{S}^c} \Delta_{ \widetilde{S}^c} \right \rVert_1 ,
\end{align*}
where the last equality is because 
$\lVert  \Delta_S \rVert_1 = \lVert  \Delta_A \rVert_1 + \lVert  \Delta_{ S \cap \widetilde{S}} \rVert_1 $.

In addition, 
\begin{align*}
-\frac{1}{T} \int_0^{T}  \left( \nabla\ell(t;\bm{\theta})  \right)^\top \Delta  \le 
\left \lVert -\frac{1}{T} \int_0^{T}  \nabla\ell(t;\bm{\theta}) \right \rVert_\infty \lVert \Delta \rVert_1 . 
\end{align*}
%\textcolor{red}{ 
Taking $\rho_1 = \frac{1}{\sqrt{M}} \rho_2 = 2\left \lVert -\frac{1}{T} \int_0^{T}  \nabla\ell(t;\bm{\theta}) \right  \rVert_\infty$ ,
%}
we get 
\begin{align*}
\frac{1}{\sqrt{M}}\lVert \Delta_{S^c } \rVert_1 + 2 \lVert  D_{., \widetilde{S}^c} \Delta_{ \widetilde{S}^c }\rVert_1
\le \frac{3}{\sqrt{M}} \lVert \Delta_S \rVert_1
+ 2  \lVert  D_{.,\widetilde{S} } \Delta_{ \widetilde{S}} \rVert_1. 
\end{align*}
Let 
$$
\mathcal{C}  =
\left \{ 
\Delta \in R^{M(p+1)} : 
\frac{1}{\sqrt{M}}\lVert \Delta_{S^c } \rVert_1 + 2 \lVert  D_{., \widetilde{S}^c} \Delta_{ \widetilde{S}^c }\rVert_1
\le \frac{3}{\sqrt{M}} \lVert \Delta_S \rVert_1 
+ 2  \lVert  D_{.,\widetilde{S} } \Delta_{ \widetilde{S}} \rVert_1 
\right \}.
$$
By Condition~\ref{def:RSC}, $\forall \Delta \in \mathcal{C}$,
there exists $\eta , c, C >0 $ such that
\begin{align}
\label{eq:RSC}
\min_{\Delta \in \mathcal{C}} 
\Delta^\top \left( \frac{1}{T}\int_0^T \nabla^2\ell(t;\bm{\theta}) \right) \Delta   \ge  \eta \lVert \Delta \rVert^2_2, 
\end{align}
with probability at least $1-c p^2  \sum_{m=1}^M T_m \exp(- C T_m^{1/5}) $.

Moreover, letting $\Delta^{(m)} =\widehat{\bm{\theta}}^{(m)} - \bm{\theta}^{(m)}$ and  $\Delta^{(m)}_j = \widehat{\theta}^{(m)}_j - \theta^{(m)}_j $, 
\begin{align}
\label{eq:bound_D} 
\lVert D_{.,\widetilde{S}}\Delta_{\widetilde{S}} \rVert_1  
&= \sum_{ 
\substack{ m\ne m' \in \widetilde{S} } } w_{m,m'} 
\left \lVert  \Delta_j^{(m)} - \Delta_j^{(m')}  \right \rVert_1  \nonumber \\
&\le  
 \sum_{ 
\substack{ m\ne m' \in \widetilde{S} } } 
w_{m,m'} \left ( 
\left \lVert   \Delta^{(m)}_j \rVert_1   + \lVert \Delta^{(m')}_j \right \rVert_1 \right)  \nonumber  \\
% &\le 
% \frac{1}{2}\sum_{m  \in \widetilde{S}} \sum_{m' \ne m  \in \widetilde{S}} w_{m,m'}  \lVert  \Delta_{m} \rVert_1  +
% \frac{1}{2}\sum_{m'  \in \widetilde{S}} \sum_{m \ne m'  \in \widetilde{S}} w_{m,m'}  \lVert  \Delta_{m'} \rVert_1
& = 
\sum_{ 
m \in \widetilde{S}} 
\sum_{\substack{ m' \in \widetilde{S} \\m'\ne m  } } 
w_{m,m'}  \left \lVert  \Delta^{(m)}_j  \right \rVert_1  \nonumber \\
%
% &= 
% \sum_{ 
% (j,m) \in \widetilde{S}} 
% \lVert  \Delta^{(m)}_j \rVert_1  
% \sum_{ m'\ne m \in \widetilde{S} } w_{m,m'}  \nonumber \\
% &= \sum_{ 
% (j,m) \in \widetilde{S}} 
% \lVert  \Delta^{(m)}_j \rVert_1  
% \phi^{(m)}_{\widetilde{S}}   \nonumber \\
% &=  \sum_{m \in \widetilde{S}} \phi^{(m)}_{\widetilde{S}}   \sum_{ 
% (j,m) \in \widetilde{S}^{(m)}} 
% \lVert  \Delta^{(m)}_j \rVert_1  
%  \nonumber \\
%  &=  \sum_{m \in \widetilde{S}} \phi^{(m)}_{\widetilde{S}}   
% \lVert  \Delta^{(m)}_{\widetilde{S}^{(m)}} \rVert_1  
%  \nonumber \\
%
&\le \phi_{\widetilde{S}} 
 \left \lVert  \Delta_{\widetilde{S} } \right \rVert_1, 
\end{align}
where $\phi_{\widetilde{S}} = \max_{m  \in \widetilde{S}} \sum_{m' \ne m  \in \widetilde{S}} w_{m,m'}$. Here, with a little abuse of notation, $m \in  \widetilde{S}$ means there exists $j$ such that $(j,m) \in \widetilde{S}$.
Because the weights are normalized---i.e., $\sum \limits_{1\le m\ne m' \le M} w_{m,m'}= 1$, 
$ 
\phi_{\widetilde{S}}    \le \max_{1\le m \le M} \sum_{1 \le m' \ne m \le M} w_{m,m'}    \le 1 
$.
% when the weights are unknown, we always have
% $  \lVert D_{.,\widetilde{S}}\Delta_{\widetilde{S}} \rVert_1 
%     \le \lVert  \Delta_{\widetilde{S} } \rVert_1 $.  
%

Next, plugging \eqref{eq:RSC} and \eqref{eq:bound_D} into \eqref{eq:basic_ineq},  
\begin{align}
\eta \lVert \Delta \rVert^2_2  
&\le 3 \frac{\rho_2  }{\sqrt{M} }  \lVert \Delta_S \rVert_1 
   - \frac{\rho_2  }{\sqrt{M} } \lVert \Delta_{S^c}\rVert_1
  + 2  \rho_2\lVert D_{.,\widetilde{S}  }\Delta_{\widetilde{S}  }\rVert_1 
  - 2  \rho_2 \lVert D_{., \widetilde{S} ^c}\Delta_{ \widetilde{S} ^c}\rVert_1 \nonumber \\
%   &\le 3 \frac{\rho_2  }{\sqrt{M} }  \lVert \Delta_{A \backslash \widetilde{S}} \rVert_1 
%   + 
%   3 \frac{\rho_2  }{\sqrt{M} }  \lVert \Delta_{ \widetilde{S}  } \rVert_1  
%   + 2  \rho_2\lVert D_{.,\widetilde{S}  }\Delta_{\widetilde{S}  }\rVert_1  \nonumber \\
&\le 
3  \frac{\rho_2\sqrt{d^* M}   }{ \sqrt{M} }  \lVert \Delta_S  \rVert_2
% + 
% 3  \frac{\rho_2\sqrt{r^*}   }{ \sqrt{M} }  \lVert \Delta_{\widetilde{S}}   \rVert_2 
+ 2   \rho_2 \phi_{\widetilde{S}} \sqrt{r^*}\lVert \Delta_{\widetilde{S} }  \rVert_2 \nonumber 
\\
&\le \rho_2 \left(3\sqrt{d^*} +  2  \phi_{\widetilde{S}} \sqrt{r^*}  \right) \lVert \Delta \rVert_2 
\label{eq:err_bound}
,
\end{align}
where the second inequality follows from  
$ \lVert \Delta_S \rVert_1 \le
\sqrt{ |S| }\lVert \Delta_S \rVert_2 $, and $  |S| \le  d^* M$
, and
 $ \lVert \Delta_{\widetilde{S} } \rVert_1 \le
 \sqrt{ |\widetilde{S}| }\lVert\Delta_{\widetilde{S} }  \rVert_2 $ 
and $|\widetilde{S}| \le r^* $. %, and the last inequality is by $ r^* \le d^* M$. 

%\textcolor{red}{
Finally, we reach the desired conclusion by plugging $\rho_2  =  2 \sqrt{M} \left \lVert -\frac{1}{T} \int_0^{T}  \nabla\ell(t;\bm{\theta}) \right  \rVert_\infty$ in \eqref{eq:err_bound}, and by Condition~\ref{def:tail_bound}, 
with probability at least $1- c' p M\exp(-T^{1/5}M^{-1} )$,
	\begin{align*}
  \left \lVert -\frac{1}{T} \int_0^{T}  \nabla\ell(t;\bm{\theta}) \right  \rVert_\infty
	\le C' M^{-1/2}T^{-2/5} , 
	\end{align*}
where $c', C'$ are positive constants. \hfill$\qedsymbol{}$
%}
%
%The final conclusion is achieved by taking a union bound over $i \in \{1,\dots, p^2 \}$.\\
\\
\\
\noindent
\textbf{Proof of Corollary~\ref{corollary1}}:
We first verify the conditions for the linear Hawkes model with least square loss --- i.e., $\ell(a,b) = (a-b)^2 $. In this case, we have 
\begin{align*}
\nabla\ell(t;\bm{\theta}) 
&=  2 \left(  dN(t) - \lambda(t)dt \right) \bm{x}(t) .
\end{align*}
By Lemma~\ref{lemma_vandergeer1995} and taking the union bound over all entries of $\bm{x}(t)$, 
\begin{align*}
\left \lVert \frac{1}{T} \int_0^{T} \nabla\ell(t;\bm{\theta})  \right \rVert_\infty 
= 
\left \lVert
\frac{1}{T} \int_0^{T}2\left(  dN(t) - \lambda(t) dt  \right) \bm{x}(t)  \right \rVert_\infty   \le C M^{-1/2} T^{-2/5}   ,
\end{align*}
with probability at least $1 - CpM\exp(- M^{-1}T^{1/5})$. Thus, Condition~\ref{def:tail_bound} is satisfied.
 
In addition,  
\begin{align*} 
\frac{1}{T}\int_0^T \nabla^2 \ell(t;\bm{\theta})  &=   \frac{1}{T} \int_0^{T}    \bm{x}(t)  
 \bm{x}^\top(t)  dt .
\end{align*}
Thus, Condition~\ref{def:RSC} is satisfied after applying Lemma~\ref{lemma_min_eigen},

Next, we verify the conditions for the non-linear Hawkes process with the exponential-link function, $g(\cdot) = \exp(\cdot)$ and estimated using the negative log likelihood loss -i.e. 
$\ell \left ( a,  b \right )  = - a \log (b)  + b $. In this case, we have 
\begin{align*}
\nabla\ell(t;\bm{\theta}) 
&=    \left(  dN(t) - \lambda(t)dt \right) \bm{x}(t)  .
\end{align*}
Similar to the linear case, Condition~\ref{def:tail_bound} is satisfied using Lemma~\ref{lemma_vandergeer1995}.
 
Under Assumption~\ref{assumption2}, there exists $\lambda_{\min}$ such that $\lambda^{(m)}(t) \ge \lambda_{\min} > 0$
\begin{align*} 
\frac{1}{T} \int_0^T  \nabla^2 \ell(t;\bm{\theta}) 
= \frac{1}{T} \int_0^T  \lambda(t)     \bm{x}(t)  
 \bm{x}^\top(t)    dt  
\ge \lambda_{\min} \frac{1}{T} \int_0^{T}    \bm{x}(t)  
 \bm{x}^\top(t)    dt  .
\end{align*}
Thus, Condition~\ref{def:RSC} is satisfied following Lemma~\ref{lemma_min_eigen}. 
\hfill$\qedsymbol{}$

%\clearpage 
\hfill 

%%%%%%%%%%%%%%%%%%%%%%%%%%%%%%%%%
% proof of theorem 3
%%%%%%%%%%%%%%%%%%%%%%%%%%%%%%%%%%
\noindent
\textbf{Proof of Theorem~\ref{theorem2}}:
Recall $S^{(m)} = \{ \beta^{(m)}_{ij}: \beta^{(m)}_{ij} \ne 0, 1\le i,j\le p \}$ and $S^{(m)}_C = \{ \beta^{(m)}_{ij}: \beta^{(m)}_{ij} = 0, 1\le i,j\le p \}$, $m \in \{ 1,\dots , M\}$.
To establish selection consistency, we need two parts. 
First, we show that our estimates on the true zero and non-zero coefficients can be separated with high probability; 
that is, there exists some constant $\Delta>0$ such that for $\beta_{S^{(m)}} \in S^{(m)} $ and $\beta_{S^{(m)}_C} \in S^{(m)}_C$,
$| \widehat{\beta}_{S^{(m)}} - \widehat{\beta}_{S^{(m)}_C}| \ge \Delta $ with high probability. 
By the $\beta$-min condition specified in Assumption~\ref{assumption5},
we have  $\beta^{(m)}_{ij} \in S^{(m)} \ge 2\tau$. Theorem~\ref{theorem1} shows that
for $m =1,\dots, M$ and $1\le i,j \le p$, 
$ | \widehat{\beta}^{(m)}_{ij} - \beta^{(m)}_{ij} | \le \tau $ 
with probability at least $1-c_1 p^2 M^2 T \exp(-c_2 M^{-1} T^{1/5})$. Then, 
for any $\beta_{S^{(m)}} \in S^{(m)} $ and $\beta_{S^{(m)}_C} \in S^{(m)}_C$,
\begin{align*}
| \widehat{\beta}_{S^{(m)}} - \widehat{\beta}_{S^{(m)}_C}|  
&= | \widehat{\beta}_{S^{(m)}} -\beta_{S^{(m)}} - ( \widehat{\beta}_{S^{(m)}_C} -\beta_{S^{(m)}_C}  ) + \beta_{S^{(m)}} - \beta_{S^{(m)}_C} |  \\
&\ge 
| \beta_{S^{(m)}}  - \beta_{S^{(m)}_C} | -  | \widehat{\beta}_{S^{(m)}} -\beta_{S^{(m)}} | - | \widehat{\beta}_{S^{(m)}_C} -\beta_{S^{(m)}_C} |  \\
&\ge \beta_{min} - 2\tau .
\end{align*}
This means the estimates on zero and non-zero coefficients can be separated with high probability.
% Next, we show there exists a post-selection threshold that allows to correctly identify $S^{(m)}$ and $S^{(m)}_C$ based on the estimates. In fact, the post-selection estimator is 
% $$ 
% \widetilde{\beta}  = \widehat{\beta} \mathbf{1}( |\widehat{\beta}| >  \tau) .
% $$
Next, we show that the thresholded estimator, 
$$ 
\widetilde{\beta}  = \widehat{\beta} \mathbf{1}( |\widehat{\beta}| >  \tau), 
$$
correctly identifies $S^{(m)}$ and $S^{(m)}_C$. 

By Theorem~\ref{theorem1}, 
we have $|\widehat{\beta}_{S^{(m)}_C}| \le \tau $, with probability $1-c_1 p^2M^2 T \exp(-c_2 T^{1/5})$. 
Thus,
$$ 
\widetilde{\beta}_{S^{(m)}_C} = \widehat{\beta}_{S^{(m)}_C} \mathbf{1}(\widehat{\beta}_{S^{(m)}_C} >   \tau_S) = 0,
$$ 
which means $\widetilde{\beta}$ selects $\beta_{S^{(m)}_C} $ into $S^{(m)}_C$ with high probability.
In addition, since $| \widehat{\beta}_{S^{(m)}}  -  \beta_{S^{(m)}} | \le \tau $,
$$
|  \widehat{\beta}_{S^{(m)}}  | \ge | \beta_{S^{(m)}}| - \tau \ge \beta_{min} - \tau > \tau  > 0.
$$
Therefore, 
$$ 
\widetilde{\beta}_{S^{(m)}} = \widehat{\beta}_{S^{(m)}}  \mathbf{1}(|\widehat{\beta}_{S^{(m)}}|  >   \tau) = \widehat{\beta}_{S^{(m)}} \ne 0 ,
$$
which means $\widetilde{\beta}_{S^{(m)} }$ selects $\beta_{S^{(m)}}$ into $S^{(m)}$ with high probability. 
 
Combining the two sides, the thresholded estimator $\widetilde{\beta}$ identifies $S^{(m)}$ and $S^{(m)}_C$ with high probability, for all $m=1,\dots, M$.
\hfill$\qedsymbol{}$

\hfill 

\begin{figure}[t]
\centering
\begin{minipage}[b]{.5\textwidth}
\includegraphics[width= \textwidth]{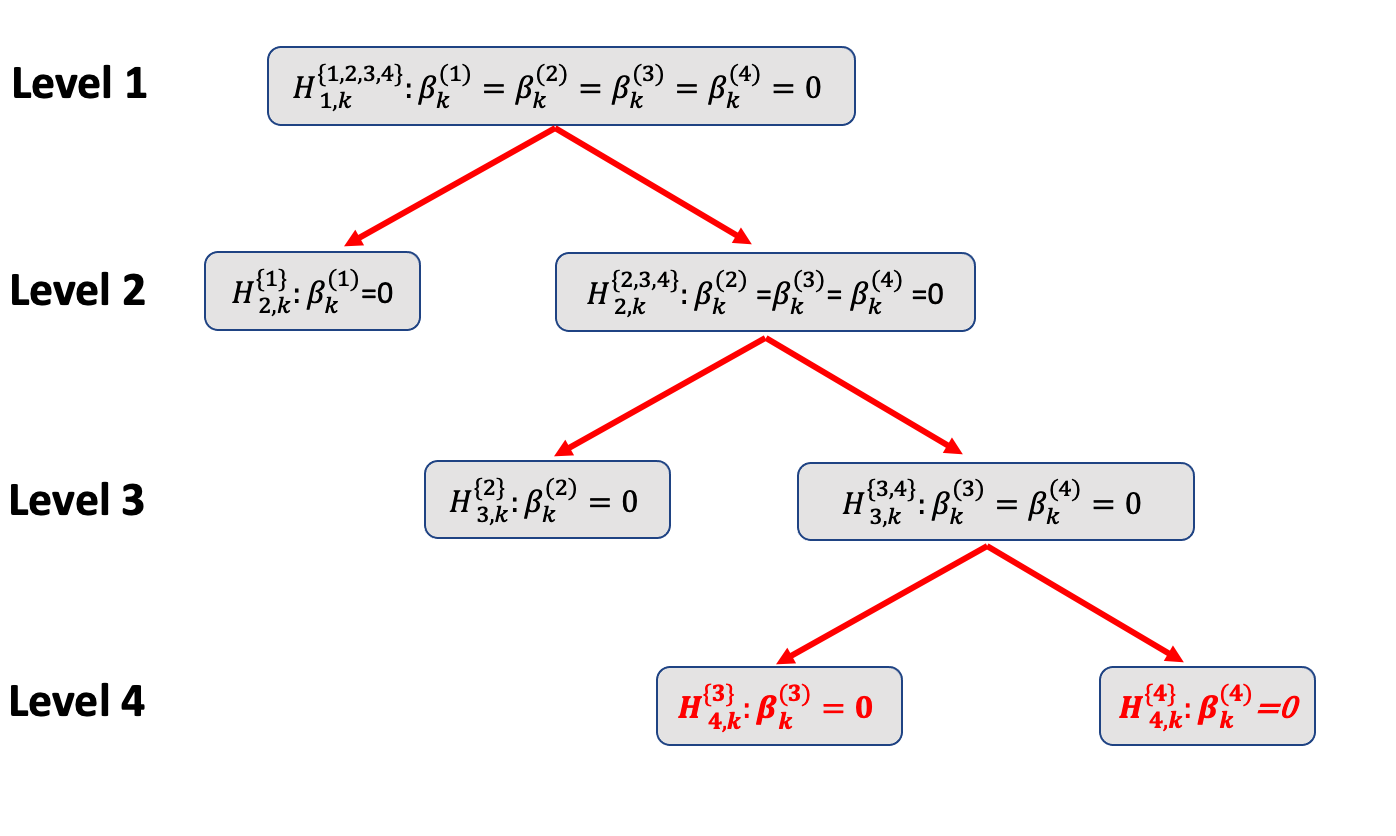}
\caption*{a} 
\end{minipage}\hfill
\begin{minipage}[b]{.5\textwidth}
\includegraphics[width=\textwidth]{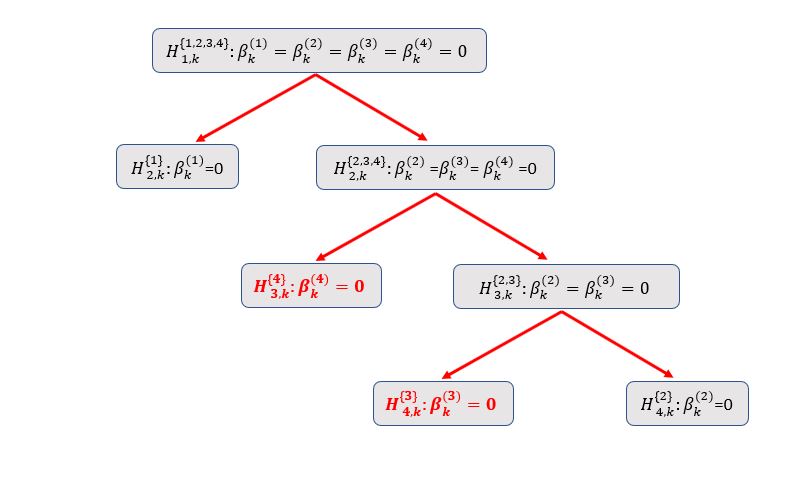}
\caption*{b}
\end{minipage}
    \caption{Two binary trees constructed to test the null hypothesis that $H_0:\beta_k^{(m)} =0$ for  $m=1,2,3,4$, where the true values of $\beta_k^{(3)}$ and  $\beta_k^{(4)}$ are 0; the leaves corresponding to these hypotheses are colored in red. The binary tree in (a) puts both zero coefficients to its bottom right branch; the binary tree in (b) swaps the position of $\beta_k^{(2)}$ and $\beta_k^{(4)}$. 
    %\as{pls make the level labels a bit larger}
    }
    \label{fig:null_tree_example}
\end{figure}

\noindent
\textbf{Proof of Theorem~\ref{fwer_oracle}}: 
% null tree first 
We start by introducing the notion of  \textit{null  tree}. 
We call a binary tree or its sub-tree a null tree if the true edge coefficients to be tested on its leaves are all zero. In any binary tree, a given zero coefficient will be associated with either (i) a single leaf associated with that coefficient; or (ii) a multi-leaf null tree, where the coefficient is tested on one of the tree's leaves. For consistency, we refer to the single leaf in (i) as a single-leaf null tree that has only this coefficient to be tested on its leaf. 
% \textcolor{red}{It follows from this definition that for any zero coefficient, there must exist a null tree with that coefficient tested on its leaf. (In a trivial case, the null tree is the leaf node itself with the zero coefficient to be tested.)} \as{Sorry, but I still don't understand this}
%\as{not clear to me} \sw{null tree is who has all leaves associated with 0 coefs/edges; a null edge is an edge with 0 coef. So if a null edge exists in the binary tree, there must be a null tree with this null edge as its leaf. For example, a sub-tree who only has a leaf node of this null edge contains this null edge}. 
The level of a null tree, $l$, is the level of its root---i.e., the length of the shortest path between the root of the null tree and the root of the binary tree plus 1. %\sw{Yes. the level concept is also general to any sub-tree.}
%\as{The null tree business seems very important yet not clear to me -- I think we may need to show this in a picture}
% see fig for illustration
As an illustration, consider a binary tree for testing the coefficients indexed $k$ in $M=4$ experiments---i.e., $\beta_k^{(1)}, \beta_k^{(2)},\beta_k^{(3)}$ and  $\beta_k^{(4)}$. Suppose $ \beta_k^{(3)} = \beta_k^{(4)} =0$. Figure~\ref{fig:null_tree_example} shows two examples of such binary trees. 
In the tree in Figure~\ref{fig:null_tree_example}a, $\beta^{(3)}_k$ and $\beta^{(4)}_k$ are associated the same two-leaf null tree of level 3, and are also associated with two separate level-4 single-leaf null trees.
In the tree in Figure~\ref{fig:null_tree_example}b, $\beta^{(3)}_k$ is associated with a single-leaf null tree of level 4 and $\beta^{(4)}_k$ is associated with a single-leaf null tree of level 3. 
% \as{is a null tree always a subset of a binary tree? if yes, we should clarify that} \sw{yes, null tree is a always a sub-tree or the tree itself (when all coefficients are 0)}
%
We call a null tree  containing a specific coefficient the \textit{largest} null tree for that coefficient if it has the highest level. 
For example, in Figure~\ref{fig:null_tree_example}a, the largest null tree for $\beta_k^{(4)}$ is the null tree of level 3. 

In the oracle binary tree, there exists a direct relationship between the level of the largest null tree and the total number of zero coefficients to be tested. 
Consider testing the coefficients indexed $k$---i.e., $\beta^{(m)}_k, m=1,\dots, M$, and suppose $r$ of these coefficients are 0. Without loss of generality, suppose $\beta^{(m)}_1= \dots =\beta^{(m)}_r = 0$. 
Now recall that the oracle binary tree puts all zero coefficients to its lower right side. 
Also, recall that by the construction of the binary tree, there are $M-l+1$ leaves to be tested under a sub-tree of level $l$.
Thus, the oracle binary tree puts all $r$ zero coefficients under a sub-tree of level $M-r+1$, meaning that the sub-tree is a level $l=M-r+1$ null tree. 
%
% Nevertheless, in the case when a binary tree constructed with arbitrary similarity, all the zero coefficients are not necessarily put to its lower right side of the tree. As a result, the number of zero coefficients to be tested on the leaves is smaller than $M-l +1$.
% \textcolor{red}{Note that when a binary tree constructed with arbitrary similarity, all the zero coefficients are not necessarily put to its lower right side of the tree. As a result, the number of zero coefficients to be tested on the leaves is smaller than $M-l +1$.} \as{this is an example of what I said above -- why do we discuss the issues with a generic tree in the proof for the oracle tree? should this come here in the proof, or perhaps in the proof of the next theorem, or even after the proof as a general comment/insight?} \sw{that makes sentence, we do not need specifically discuss a generic tree here.}
% %
% Specifically, consider a slightly misspecified binary tree where the zero coefficient on the left leaf at level $M$ of the oracle binary tree exchanges its position with the non-zero coefficient on the left leaf at level $l=M-r+ 1$. Under this misspecified binary tree, there still exists a level $l= M-r+1$ null tree (which is the left leaf at level $l=M-r+1$) but there is only 1 zero coefficient tested on its leaf. 

Next, we show that when testing all edge coefficients corresponding to the networks of $p$ nodes in $M$ experiments, our hierarchical testing procedure with the oracle binary tree controls the FWER. Throughout, we refer to the coefficients associated with edges of the networks as `nonzero coefficients' and those associated with non-edges as `zero coefficients'. %\as{shouldn't we give the proof, or at least the first part about the null tree, for a single coefficient across the M networks?}

Recall that we index the connectivity/edge coefficients from $1$ to $p^2$. Let $\mathcal{H}_{l,k}$ be the collection of null hypotheses in $H_k^{(m)}: \beta^{(m)}_k=0$, $m \in \{ 1,\dots, M\}$, where the level of their corresponding largest null-trees is $l$. Thus, for any binary tree $\mid \mathcal{H}_{l,k} \mid \le M-l + 1$, with the equality holding for the oracle binary tree when there exist at least one zero coefficient. 
%\as{something is off in this sentence and I am not really following} \sw{here $l$ is the level of the binary tree. $(m)$ indicates the conditions and $k$ indicates which edge. As discussed earlier, each $\beta^{(m)}_k=0$ must be associated with a null tree, where the null tree can be at different levels. So $\mathcal{H}_{l,k}$ indicates the $H_k^{(m)}: \beta^{(m)}_k=0$ that has their null tree at level $l$.}. 
Let $\mathcal{H}_l =\left  \{ \mathcal{H}_{l,k}: \mathcal{H}_{l,k}\ne \emptyset, 1\le k \le p^2  \right \} $; that is the collection of non-empty $\mathcal{H}_{l,k}$ sets. 
Denote $n_l = | \mathcal{H}_l |$.

Following Step~2 in the hierarchical testing procedure, the root of a level-$l$ sub-tree is rejected with probability not higher than $\alpha_l$. In addition, in our hierarchical testing procedure a lower level test is only considered when the test of its parent node is rejected. Therefore,  
\begin{align}\label{eqn:thm3pfineq}
\mathbb{P}\left( \bigcup_{ H_k^{(m)} \in \mathcal{H}_{l,k} } \text{$H_k^{(m)}$ is rejected} \right)
\le \mathbb{P}\big(\text{The root of the level-$l$ null tree is rejected} \big) \le \alpha_l.
\end{align}

%\begin{align*}
%\mathbb{P}\left( \bigcap_{ H_k^{(m)} \in %\mathcal{H}_{l,k}  } \text{$H_k^{(m)}$ is not rejected} \right)
%\ge \mathbb{P}\left( \text{ the root of the %null-tree is not rejected} \right) \le \alpha_l.
%\end{align*}

Then, the FWER is controlled as follows.
\begin{align*}
\mathrm{FWER} &= \mathbb{P}\left( \bigcup_{l=1}^M 
\bigcup_{  \mathcal{H}_{l,k} \in  \mathcal{H}_l }   
\bigcup_{ H_k^{(m)} \in \mathcal{H}_{l,k} } \text{$H_k^{(m)}$ is rejected}   
\right) \\
&\le \sum_{l=1}^M n_l \max\limits_{\mathcal{H}_{l,k}  \in  \mathcal{H}_l }
\mathbb{P}\left( \bigcup_{ H_k^{(m)} \in \mathcal{H}_{l,k} } \text{$H_k^{(m)}$ is rejected} \right) \\
&\le \sum_{l=1}^M n_l  \alpha_l, 
\end{align*}
where the first inequality is by Boole's inequality and the second inequality follows from \eqref{eqn:thm3pfineq}.

Let $\pi_0$ be the total number of zero coefficients, which is no greater than the total edges $p^2 M$. Let $C_l$ be the number of leaves under a level $l$ null-tree. 
Then, 
\begin{align*}
\sum_{l=1}^M n_l C_l = \pi_0 \le p^2 M.
\end{align*}
%By the binary tree construction, $C_l \le M-l + 1$. 
% \as{This is the only place in the actual proof that we use the null tree and I can't see how this relates to what we described above. In particular, didn't we already describe what $C_l$ would be for the oracle binary tree?} \sw{Yes, under the oracle binary tree $C_l = M-l + 1$. Originally, I wanted to first make $C_l$ general and then talk about it under the oracle tree, but it seems that is not necessary. So now I re-ordered the proof and use the $C_l$ under the oracle directly.} 
Now, recall that using an oracle binary tree, $C_l = M-l + 1$. Thus, 
taking $\alpha_l =  \frac{\alpha}{p^2}\frac{C_l}{M} =  \frac{\alpha}{p^2}\frac{M-l+1}{M} $, 
\begin{align*}
\mathrm{FWER} \le \sum_{l=1}^M  n_l \frac{\alpha}{p^2}\frac{C_l}{M} = \frac{\pi_0}{p^2 M}\alpha \le \alpha. 
\end{align*}

%\as{I really don't understand the following} \sw{ Note that the binary tree is constructed so the network of a condition with the least edges are put on the bottom right of the tree. When using the oracle similarity, we will put similar networks--i.e., also have few edges-- to the bottom right side. This way, all zero-edges are put to bottom right side under a null tree --- however, if the binary tree are not well constructed, some zero-edge will be put to the left side which makes the null tree containing fewer null edges as it should be if an oracle similarity is used. }
%\sw{need to redefine the oracle binary specific to each edge so that zero-coefficients are always put to the lower right side of the binary tree.}
%As discussed at the beginning of the proof, for the oracle binary tree, $C_l = M-l + 1$, which leads the choice of $\alpha_l = \frac{\alpha}{p^2}\frac{M-l+1}{M}$. 
\hfill$\qedsymbol{}$

\hfill 

\noindent 
\textbf{Proof of Theorem~\ref{fwer_general}}:
Next, we show that given large and sparse networks, the hierarchical testing procedure still controls the FWER for a large number of experiments without the knowledge of the oracle binary tree. 

Consider an coefficient indexed $k$ that is not zero in at least 1 experiment.
Let $r^{(k)}_l$ is the number of level-$l$ null-trees in the binary tree for the coefficient $k$. By the binary tree construction, there is at most one level $l$ null-tree, $r^{(k)}_l \le 1$. (There may be no level-$l$ null tree, in which case, $r^{(k)}_l =0$.) Then,
\begin{align*}
\sum_{l=2}^M r^{(k)}_l (M-l+1) \le  \frac{M(M-1)}{2} ,
\end{align*}
% the equality achieves when only 1 non-null edge and put to bottom right. So for the other null edges are all put to left child leaves. So r^k_l = 1 then M-1 + ... + 1 = (M-1)( M - 1 + 1) /2 
The maximum on the right-hand side of the above is achieved when there are $M-1$ zero-coefficients and one non-zero coefficient which is allocated to the bottom right leaf---i.e., the deepest level of the tree. 

Let $ {n_1}/{p^2} \le 1$ indicates the proportion of level-1 null trees --- that is, the binary trees on which the coefficients to be tested on all leaves are zero. 
%\as{??} \sw{$n_1$ are number of edges that are all zero across the $M$ conditions}
Taking $\alpha_l = \frac{\alpha}{p^2}\frac{M-l+1}{M}$,
\begin{align*}
\mathrm{FWER} &=  n_1 \alpha_1 + \sum_{l=2}^M n_l \alpha_l \\
     &=   \frac{\alpha}{p^2} n_1   + \frac{\alpha}{p^2M} \sum_{l=2}^M n_l (M-l+1) \\
     &\le   \frac{\alpha}{p^2} n_1   + \frac{\alpha}{p^2M} \frac{M(M-1)}{2} (p^2 - n_1)  %\\
  %   &= \alpha \left ( \lambda + (1-\lambda)\frac{M-1}{2}  \right ).
\end{align*}

Recall that $d^* \equiv \max_{ 1\le m\le M, 1\le i \le p} d_i^{(m)} $, where $d^{(m)}_i = |S^{(m)}_i|$ and $S^{(m)}_i= \{j:  \beta^{(m)}_{ij}\ne 0 , 1\le j \le p \}$ as defined in Section~\ref{sec:theory}.

Thus, $n_1 + d^* p M \ge p^2$, or
\begin{align*}
    p^2 - n_1 \le d^*p M  , 
\end{align*}
which leads to
\begin{align*}
\mathrm{FWER} &\le   \frac{\alpha}{p^2} n_1   + \frac{\alpha}{p^2} \frac{d^* p M (M-1)}{2} .
\end{align*}
Noting that $ n_1 \le p^2 $,  
\begin{align*}
    \mathrm{FWER} &\le \alpha \left( 1+ \frac{d^* M (M-1) }{2 p}  \right) ,
\end{align*}
as desired.
\hfill$\qedsymbol{}$

% Given that $d^* = o(p)$ and $M$ is a fixed number of experiments,
% \begin{align*}
%     FWER &\le \alpha \left( 1+  o(1) \right), 
% \end{align*}
% as $p \rightarrow \infty$. 

% given $d^* M(M-1) = o(p^2)$, we have 
% \begin{align*}
%     p^2 - n_1 \le d^* M =  o\left ( \frac{p^2}{M-1} \right) , 
% \end{align*}
% which leads  $1- \lambda = o\left ( \frac{1}{M-1} \right)$. Therefore, 
% \begin{align*}
%     FWER = \alpha\left (  1 - o\left ( \frac{1}{M-1} \right ) + o(1) \right) .
% \end{align*}

\clearpage
%%%%%%%%%%%%%%%%%%%%%%%%%%%%%%%%%%%%%%%%%%%%%%%%%%%%%
\section{Additional Simulation Results}\label{sec:addition}

\subsection{Illustration on the simulation setting in Section~\ref{sec:sims}}

\begin{figure}[h]
	\centering
	\includegraphics[width=\linewidth, clip=TRUE, trim=0mm 10mm 0mm 0mm]{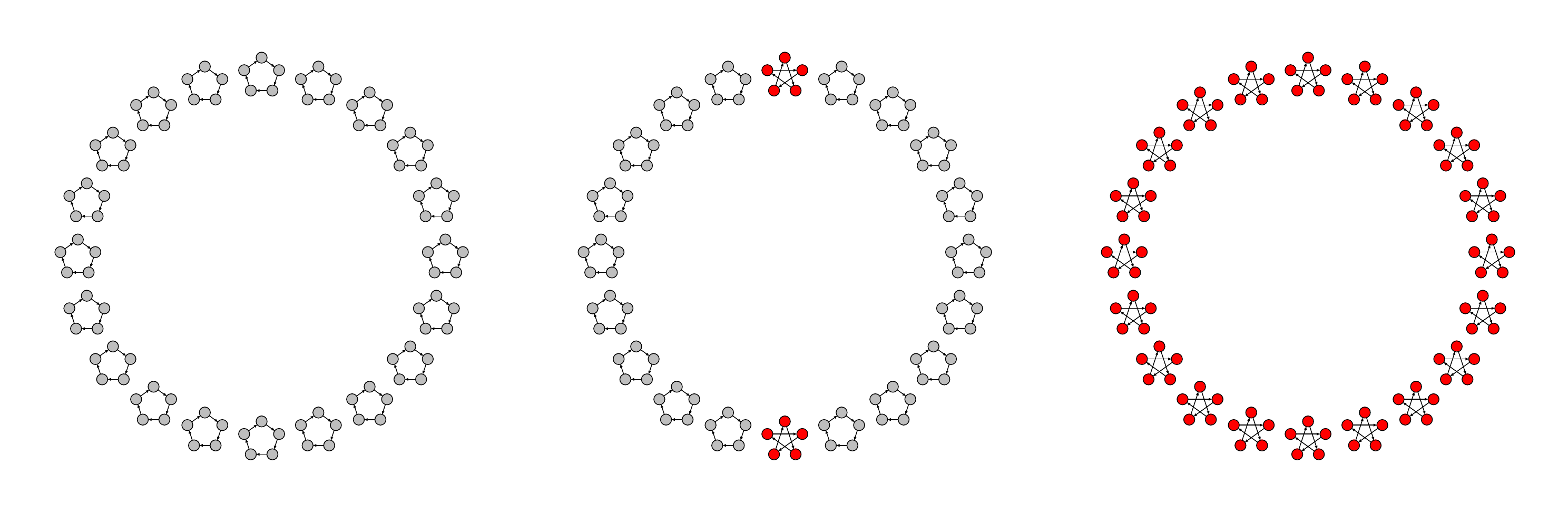}
	\caption{Networks of $p=100$ processes under $M=3$ experiments. Network~1 (left) consists of 20 circles, Network~3 (right) consists of 20 stars, and Network~2 (middle) is a mix of 18 circles and 2 stars. }
	\label{fig:network_star_circle}
\end{figure}

In Section~\ref{sec:sims}, we consider $M=3$ networks of $p=100$ linear Hawkes processes. The networks are designed such that Networks~1 and 2 are much more similar to each other than Network~3. Specifically, Network~1 and 3 consists of 20 5-node  circles and stars, respectively, and Network~2 is a mix of 18 circles and 2 stars (see Figure~\ref{fig:network_star_circle}).

\subsection{Hierarchical testing with incorrect hierarchy}
% \as{what does active and inactive mean here?} \sw{moderately sparse and highly sparse...}
To illustrate how a poorly constructed hierarchy, i.e., binary tree, affects the power of the hierarchical testing procedure, we consider networks of $p=100$ nodes under $M=1,5,10, 20, 30, 50$ experiments. Half of the networks are set to be highly sparse with 0.1\% edges. The other half are moderately sparse (referred to as `non-sparse' in the following) with 5\% edges. Sparse and non-sparse networks do not share any common edges. %\sw{confirmed}
%Note that the oracle binary tree \as{is required to} put all sparse networks to deeper levels of the tree. 

The poorly constructed binary tree we consider assigns the coefficients associated with non-sparse networks to deeper levels of the tree, resulting in nonzero coefficients at the deeper levels of the tree. This is in contrast to the oracle binary tree, which always assigns the zero coefficients to the deeper levels.

\begin{figure}[!t]
	\centering
	\includegraphics[width=1\linewidth]{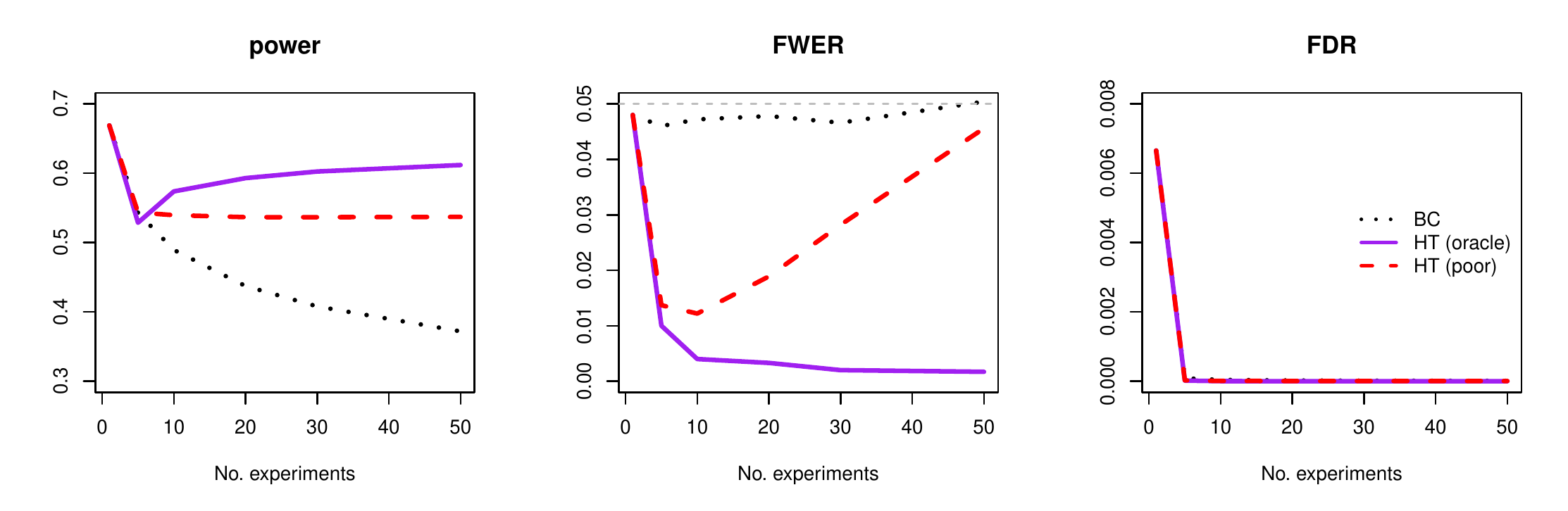}
	\caption{Power, FWER and false discovery rate (FDR) between Bonferroni correction (BC) and the hierarchical testing procedure using the oracle and poorly constructed binary trees (poor). The FWER is controlled at $\alpha =0.05$ (gray dashline). %\as{same as the main paper}
}
	\label{fig:HT_plot2}
\end{figure}

As expected, when using a poorly constructed hierarchy, the power of our procedure deteriorates compared with the oracle binary tree. However, the procedure still controls the FWER 
%\as{I just realized that we may be using `tightly' incorrectly throughout -- it seems that we actually control the FWER conservatively -- what do you mean by `tightly'?} \sw{you are right! by 'tightly' I actually mean 'conservatively' because our result shows the FWER is upper bounded by alpha, not exactly alpha} 
and is more powerful than Bonferroni correction (see Figure~\ref{fig:HT_plot2}).  

% we see an increasing trend in FWER under the poorly constructed dendrogram. This empirical result is consistent with our theoretical finding where the upper bound of the FWER is increasing as the number of experiment goes up (see the proof in Theorem~\ref{chap3:fwer_general}. 

\subsection{Effect of increased number of experiments}
In this simulation, we investigate the benefits of the proposed estimation procedure as the number of experiments increases. To this end, we consider networks under four conditions, where the first three conditions are exactly the same as in Figure~\ref{fig:network_star_circle} and the fourth network has the same structure as Network~1. The experiment lengths for the first three condition are the same as those in Section~\ref{sec:sims-est} and 500 in the fourth condition; that is $T_m = 200, 500, 300, 500$ for $m=1,2,3,4$, respectively.  

\begin{figure}[t!]
\centering
\subfloat[]{\includegraphics[width=0.5\linewidth]{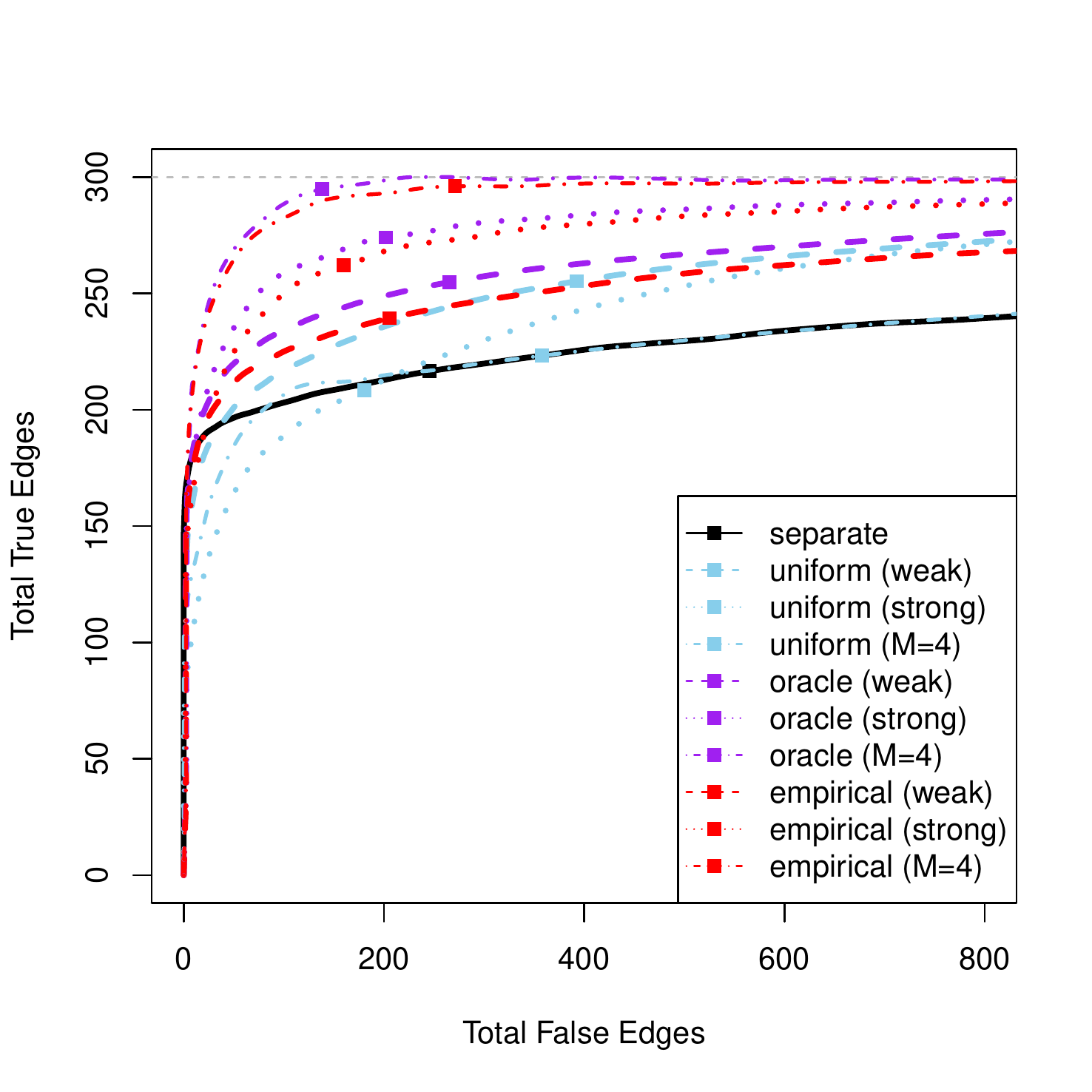}} 
\quad 
\subfloat[]{\includegraphics[width=0.415\linewidth]{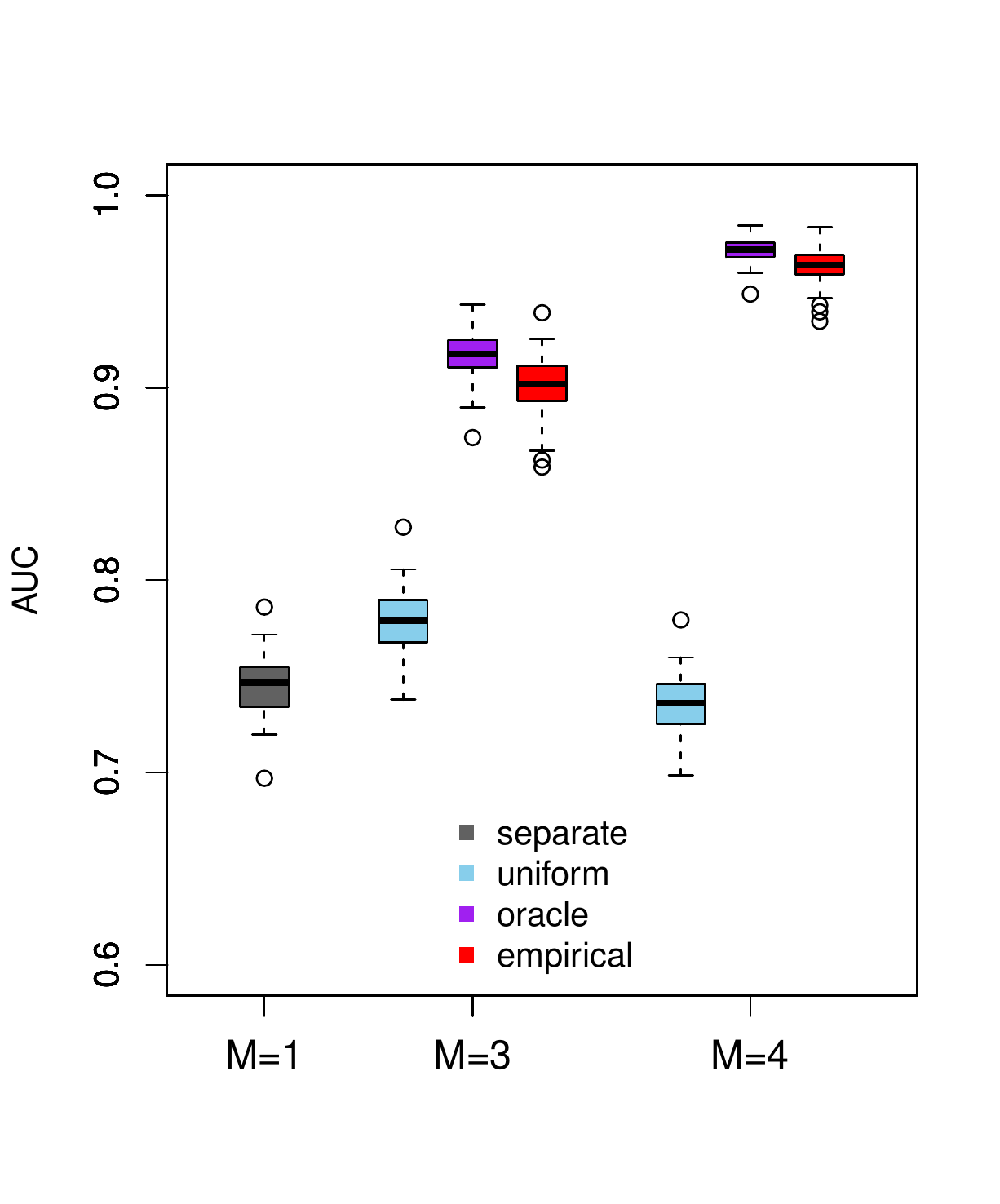}}
	\caption{Edge selection performance of the proposed joint estimation method in a simulation study evaluating the benefit of including extra experiments in the estimation procedure. The main result focused on inferring edges using 3 networks of linear Hawkes processes as in Figure~\ref{fig:network_star_circle}. The performance is compared with the result using 4 networks (indicated as $M=4$) with strong fusion penalty where the first three is the same as before and the extra one has the same structure as the first network. The plots in (a) show average number of true positive and false positive edges, over 100 simulation runs, for the joint estimation method with different choices of weights, compared to separate estimation of each network. Weight strategies include oracle, empirical and uniform weights. Solid squares ($\blacksquare$) correspond the choice of tuning parameter using eBIC. 
	The boxplots in (b) show the distribution of the area under the curve (AUC) values corresponding to the edge selection performance in (a) over 100 simulation runs for separate estimation ($M=1$), and the joint estimation with different choices of weights using $M=3$ and $M=4$ experiments. The total numbers of true and false edges are normalized between 0 and 1 when calculating the AUCs.
	}
		\label{fig:lstpfp3}
\end{figure}

The results, summarized in Figure~\ref{fig:lstpfp3}, show that the edge selection performance of the proposed methods improves as the number of experiments increases. More specifically, with informative weights (either oracle or empirical), the area under the true positive false positive curves (AUC) improves as the number of experiments increases, whereas the performance deteriorates when noninformative (uniform) weights are used. 
This finding corroborates our theoretical results, where the error bound becomes tighter when the total experiment length gets larger and more similar conditions are involved.

\end{document}